\documentclass[twocolumn,showpacs,showkeys,preprintnumbers,amsmath,amssymb]{revtex4}

\usepackage{graphicx}
\usepackage{dcolumn}
\usepackage{bm}
\usepackage{hyperref}
\usepackage{color}

\begin{document}

\preprint{IRE \& NSC "KIPT" of NASU/v.2-Ubitron}

\title{Hybrid planar FEM in magnetoresonance regime:
       control of dynamical chaos}

\author{Vitaliy A. Goryashko}
\author{Kostyantyn Ilyenko}%
 \email{kost@ire.kharkov.ua}
\affiliation{%
Institute for Radiophysics and Electronics of NAS of Ukraine \\ 12
Acad. Proskura Street, Kharkiv, 61085, Ukraine}%

\author{Anatoliy Opanasenko}
\affiliation{ National Science Center "Kharkiv Institute of
Physics and Technology" of NAS of Ukraine \\ 1 Akademichna Street,
Kharkiv, 61108, Ukraine
}%

\date{June 18, 2008}

\begin{abstract}
We establish the influence of nonlinear electron dynamics in
the magnetostatic field of a hybrid planar free-electron maser
on its gain and interaction efficiency. Even for the `ideal'
undulator magnetic field the presence of uniform longitudinal (guide)
magnetic field potentially leads to the existence of chaotic zone
around certain (magnetoresonant) value of the guide magnetic field.
The width of the chaotic zone is given by the Chirikov resonance-overlap
criterion applied to the normal undulator and cyclotron
frequencies with respect to the coupling induced by the
undulator magnetic field. Using analytical asymptotically exact
solutions for trajectories of individual test electrons, we show
that the magnetoresonant multiplier in electron trajectories
is also present in the expression for the gain. The same Chirikov
resonance-overlap criterion allows us to estimate analytically the
maximal magnetoresonant gain of a hybrid planar free-electron
maser showing that, in spite of the well-known drop in the gain
for the exact magnetoresonance, the operation regime in the zone
of regular dynamics slightly above the magnetoresonant value of
the guide magnetic field is the preferable one.
\end{abstract}

\maketitle

\section{\label{sec:Intro}%
Introduction}%

Starting with the successful
experiment~\cite{SprangleGranatstein1978,GranatsteinEtAl1982},
previous three decades have witnessed a spectacular development of
theory and experiment of free-electron maser with guide (uniform
longitudinal) magnetic field (hybrid FEM). Utilizing the Doppler
frequency upshift, free-electron masers and lasers posses a unique
property to amplify and generate coherent electromagnetic
radiation across nearly complete electromagnetic spectrum: from
radio waves to vacuum ultraviolet~\cite{VUV2000}. An interest to the
hybrid FEM is reinforced by a promise to attain a substantial
microwave power level (up to a few gigawatts) in their planar
configuration because of the use of superwide sheet electron beams
(e.g., in~\cite{ArzhannikovEtAl1998} emission and transport of a
140~cm-wide, 20~kA and 2~MeV sheet electron beam for purported FEM
applications was reported). Sheet electron beams allow one to
weaken restrictions on the maximal aggregate beam current posed by
space charge effects while reaching the current values of about
30~$\div$~50~kA. This makes the hybrid planar FEM one of the most
attractive sources of powerful electromagnetic radiation in the
Terahertz Gap: the frequency range from 0.3~to 3~THz.

In pioneering works~\cite{Colson1979,Freund1981,KokhmanKulish1984,
Marshall1986,GinzburgNovozhilova1986} analytical investigations of
stationary regimes of microwave amplification and generation in a
hybrid FEM were carried out; later these results were refined
mainly through numerical simulations.
Experimentalists~\cite{CondeBekefi1991,KaminskyEtAl1991,Japanese1993,NGorky1996}
reported a considerable loss of electron beam current and microwave
power for a hybrid FEM for a certain range of values of
the guide magnetic field. A study of chaotic particle
dynamics in free-electron lasers was undertaken in
papers~\cite{ChenDavidson1990,ChenDavidson1991}, where the authors
investigated effects of high-current (high-density) regime and the
transverse spatial gradients in the applied wiggler magnetostatic
field. Nevertheless, to the present day there still exist some
fundamental questions, which either have no answers or only
partial ones (cf.~\cite[pp.~430-439]{FreundAntonsen1995}): under
what conditions motion of an individual test electron becomes
chaotic in the presence of guide magnetic field; what does
influence the width of dynamical chaos zone around the
magnetoresonant value (the undulator frequency is approximately
equal to the cyclotron frequency) of the guide magnetic field;
which maximal portion of initial kinetic energy from the
longitudinal motion can be transferred to the
transversally-vibrational motion (in an FEM the transfer of beam
energy from longitudinal constant motion  to the microwave field
takes place indirectly through coupling of transversally-vibrational
degrees of freedom to the microwave); does such an interaction with
the microwave cause any widening of the chaotic dynamics zone; and,
finally, what does define the maximal value of the gain under the
conditions of the magnetoresonance? 

We focus our attention on the operation of FEM around the
magnetoresonant regime caused by the guide magnetic field
usually used in FEM setups to enhance the efficiency of
beam-microwave interaction and provide transverse confinement of
electron beam. The study is based upon analytical asymptotically
exact expressions for electron trajectories developed
recently~\cite{US2004,US2006} as well as on the use of direct
numerical simulations. More specifically, in the adopted approach
we highlight the fundamental role of nonlinear dynamical system
describing motion of electrons in the combined magnetostatic spatially
periodic (undulator) and uniform guide magnetic field.

The next Section contains nonlinear and linearized in the
microwave field self-consistent systems of equations governing a
hybrid planar FEM amplifier. In Section~\ref{sec:MR}, we find
asymptotically exact solutions of equations of motion of an
individual test electron in the magnetostatic field of a
hybrid planar FEM and study their properties.
Section~\ref{sec:LTMA} contains derivation of dispersion equations
and calculation of magnetoresonant growth rates. A comparison with
direct numerical simulations of self-consistent nonlinear system
of governing equations is presented in Section~\ref{sec:NLS}.
The article ends with the Summary and Discussions Section.

\section{\label{sec:GE}Governing Equations}

\begin{figure}[b]
\includegraphics[221,590][335,704]{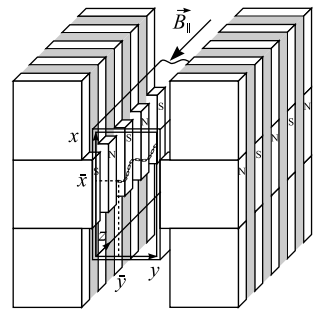}
\caption{\label{fig:ProblemGeometry} Schematic drawing of rectangular
waveguide with mounted permanent magnet planar undulator immersed
in the uniform longitudinal (guide) magnetic field.}
\end{figure}

We regard that at the entrance of the interaction region
(cross-section $z=0$) of a hybrid planar FEM the electron beam is
continuous and unmodulated. In the amplifier regime there also is a
seed microwave signal. It can be shown that for a regular
electrodynamics structure the microwave field and electron beam
current density in the interaction region are periodic functions of
time. Neglecting temporal harmonics generation, we will assume that there
occurs resonant interaction only with one spatial
harmonic of the microwave field. Under such assumptions the
microwave field in an ideally conducting regular waveguide with the
electron beam
\begin{equation}\label{single_mode}
\begin{split}
 \vec{E}(\vec{r},\!t) & \!\!\approx\!  \mathrm{Re}\{C\vec{E}^{0}(\vec{r},\!t)\}\!\! =\!
 \mathrm{Re}\{C(z)\vec{e}(\vec{r}_{\perp})e^{-i(\omega t - k^{0}_{z}z)}\},  \\
 \vec{B}(\vec{r},\!t) & \!\!\approx\!  \mathrm{Re}\{C\vec{B}^{0}(\vec{r},\!t)\}\!\! =\!
 \mathrm{Re}\{C(z)\vec{b}(\vec{r}_{\perp}) e^{-i(\omega t - k^{0}_{z}z)}\}
\end{split}
\end{equation}
differs from that of without the electron beam only by an
amplitude multiplier $C(z)$ ($\vec{E}^{0}(\vec{r},t)$ and
$\vec{B}^{0}(\vec{r},t)$ obey homogeneous Maxwell's equation);
$\vec{e}(\vec{r}_{\perp})$ and $\vec{b}(\vec{r}_{\perp})$ are the
membrane eigenfunctions of the coordinates $\vec{r}_{\perp}=
(x,y)$ in the waveguide cross-section $S$ (microwaves and the
electron beam propagate in the positive direction of the $z$-axis,
see Fig.~\ref{fig:ProblemGeometry}); $k^{0}_{z}$ is the `cold'
propagation constant of the eigenwave.

In the case of low-density electron beam, one can neglect its
space-charge and, by averaging on the electrons' entrance phase,
reduce the problem of interaction of electron beam with a seed
synchronous microwave of frequency $\omega$ to the task of
solving a single-particle equations of motion for arbitrary
entrance time $t_{e}$ and Maxwell's equations in the form of
Kisunko-Vainshtein's equations of excitation for regular
waveguides~\cite{Kisunko1949,VanSoln1973}. It is exactly in
this manner a highly successful theory of the gyrotron was
initially developed by Gaponov~\cite{Gaponov1961}.

Following \cite[p.~200]{VanSoln1973} (see also~\cite{Gover1995}),
let us write the self-consistent nonlinear system of equations in
the form:
\begin{subequations}\label{Master_Eqns}
\begin{eqnarray}
m_e\frac{d(\gamma\vec{v})}{dt} & = & e\vec{E} +
\frac{e}{c}[\vec{v}\times (\vec{B}_{p} + \vec{B})],
\label{Eqns_Motion} \\
\frac{dC}{dz} & = & -\frac{\omega}{2\pi
P_{0}}\int\limits_{S}\!\!\!\int\limits_{-\pi /\omega}^{\pi
/\omega}\!\!\!\!
(\vec{j}\cdot(\vec{E}^{0})%
\hspace{-.5 mm}\raisebox{1.ex}[0pt][0pt]{\scriptsize
\ensuremath{\ast}})e^{i\omega t}dtdS, \label{Eqns_Excitation}
\end{eqnarray}
\end{subequations}
where
\begin{eqnarray}\label{Notations01}
\vec j(\vec r, t) & = & e\!\!\sum\limits_{k}
\delta(\vec{r}-\vec{r}^{\,k})\vec{v}^{\,k}, \nonumber \\
P_{0} & = &
\frac{c}{8\pi}\mbox{Re}\int_{S}([\vec{e}(\vec{r}_{\perp})\times
\vec{b}^{\ast}(\vec{r}_{\perp})]\cdot\vec{e}_{z})dS. \nonumber
\end{eqnarray}
Here $m_{e}$ and $e$ are the electron rest mass and charge,
respectively; $c$ is the speed of light; $\vec{v}=d\vec{r}/dt$,
$\gamma = (1 - \vec{v}\,^{2}/c^{2})^{-1/2}$ is the relativistic
factor; $\vec{j}(\vec{r},t)$ is the electron current density;
$\vec{r}^{\,k}(t) = \vec{r}(t,t^k_e)$ and $\vec{v}^{\,k}(t) =
\vec{v}(t,t^k_e)$ are the coordinates and velocity of a $k$'s
electron, which entered the interaction region at the moment of
time $t^k_e$; the star (${}^{*}$) denotes complex conjugation;
$\vec{e}_{z}$ is the unit vector along the $z$-axis; the overbar
in equation~\eqref{Eqns_Excitation} stands for the time average,
which is effectively reduced to the averaging over the electrons'
entrance phase. Initial conditions for system~\eqref{Master_Eqns}
consist in the specification of coordinates, velocities and the
amplitude of microwave field at the entrance to the interaction region:
$\vec{r}\,(t_{e},t_{e})=(\bar{x},\bar{y}, 0)$,
$\vec{v}\,(t_{e},t_{e})=(V_{x},V_{y},V_{\|})$ and $C(0) = C_{0}$.

Magnetostatic field $\vec{B}_{p}(\vec{r})$
consists of a guide (uniform longitudinal) magnetic field
$\vec{B}_{\|}$ and planar
spatially periodic undulator magnetic field
$\vec{B}_{w}(\vec{r})=$ $-B_{\perp}(0, \cosh(2\pi [y -
b'/2]/\lambda_{w})\sin(2\pi z/\lambda_{w}), \sinh(2\pi [y -
b'/2]/\lambda_{w})\cos(2\pi z/\lambda_{w}))$. Having the primary
goal in demonstration of underlying physics, we intentionally consider
a simplest possible model of a hybrid planar FEM (which accounts for the
undulator through only one component of its spatially periodic magnetic
field taken in the limit $2\pi [y -
b'/2]/\lambda_{w}$ $\ll$ $1$):
\begin{equation}\label{HybPump_Field}
\vec{B}_{p}=(0,-B_{\perp}\sin(2\pi z/\lambda_{w}),-B_{\|}),
\end{equation}
where $B_{\perp}$ is a constant amplitude
\cite{Japanese1993,Destler1996} and $b'$ is the undulator gap
width. However, the developed approach can be extended
straightforwardly on `realisable' undulator/wiggler magnetic fields
and space charge dominated beams.

In order to explore analytically the linear stationary regime of
amplification, we linearize
equations~\eqref{Master_Eqns} in the microwave fields $\vec{E}$
and $\vec{B}$ ($B_{\|}$, $B_{\perp}$ $\gg$ $c|\vec{E}|/|\vec{v}|$,
$|\vec{B}|$). Trajectory of an electron is primarily given by its
motion in the magnetostatic field
\begin{equation*}\label{Trajectory}
\vec{r}(t,t_{e}) \equiv \vec{r}(\tilde{t},t_{e}) =
\vec{r}_{0}(\tilde{t}) + \mathrm{Re}\{\vec{r}_{1}(\tilde{t})e^{i
\omega t_{e}}\} \,\,\,\,(|\vec{r}_{0}| \gg |\vec{r}_{1}|);
\end{equation*}
$\tilde{t} = t - t_{e}$ stands for the transit time of an electron.
We also hold that $C(z)$ is a slow function of $z$. In the
zeroth-order in the microwave field approximation
system~\eqref{Master_Eqns} reduces to equations of motion in the
magnetostatic field
\begin{equation}\label{eqn:ZO_Eqns_Motion}
m_e\frac{d(\gamma_{0}\vec{v}_{0})}{d\tilde{t}} =
\frac{e}{c}\,[\vec{v}_{0}\times\vec{B}_{p}(\vec{r}_{0})],
\end{equation}
where $\vec{v}_{0}=d\vec{r}_{0}/d\tilde{t}$ and $\gamma_{0}=(1 -
\vec{v}_{0}^{\,2}/c^2)^{-1/2}$. It then immediately follows that
in this approximation the total kinetic energy is conserved (i.e.
$\gamma_{0}=$~const; $\vec{v}_{0}^{\,2}=\vec{v}^{\,2}(t_{e},t_{e})
\equiv V_{x}^{2} + V_{y}^{2} + V_{\|}^{2}$), which greatly
simplifies the analysis of system \eqref{eqn:ZO_Eqns_Motion}. It
should be emphasized that properties of solutions to equations
\eqref{eqn:ZO_Eqns_Motion} are fundamental to the consideration of
microwave amplification and generation in a
FEM~\cite{FriedHirsh1980,Colson1981,Freund1981}. In the first
order (linear) approximation in the microwave field, we obtain a
coupled system of equations for $\vec{r}_{1}$ and linearized
equations of excitation
\begin{subequations}\label{FO_System}
\begin{eqnarray}
m_e\gamma_{0}\frac{d}{d\tilde{t}}\Big[\vec{v}_{1} & + &
\gamma_{0}^{2}\vec{v}_{0}\frac{(\vec{v}_{0}\cdot\vec{v}_{1})}{c^{2}}\Big]
= eC\vec{F} + \label{FO_Eqns_Motion} \\
\frac{e}{c}\{[\vec{v}_{1} & \times & \vec{B}_{p}(\vec{r}_{0})] +
[\vec{v}_{0}\times (\vec{r}_{1}\cdot\vec{\nabla})\vec{B}_{p}
(\vec{r})|_{\vec{r}=\vec{r}_{0}}]\},
\nonumber \\
\frac{dC}{d\tilde{t}} & = & -\frac{i\omega
|I_{0}|}{4P_{0}}(\vec{r}_{1}\cdot\vec{F}^{*}),
\label{FO_Eqns_Excitation}
\end{eqnarray}
\end{subequations}
where
\begin{eqnarray}
\vec{F}(\tilde{t}) & = &
\vec{f}(\vec{r}_{0\perp},\vec{v}_{0})e^{-i(\omega
\tilde{t} - k^{0}_{z}z)}, \nonumber \\
\vec{f}(\vec{r}_{0\perp},\vec{v}_{0}) & = &
\vec{e}\,(\vec{r}_{0\perp}) + c^{-1}[\vec{v}_{0} \times
\vec{b}(\vec{r}_{0\perp})]. \nonumber
\end{eqnarray}
Here $\vec{v}_{1}=d\vec{r}_{1}/d\tilde{t}$, $I_{0}$ is the
electron beam current at the entrance to the interaction region
and $\vec{F}(\tilde{t})$ is the microwave force estimated at the
unperturbed electron trajectory. \\

\section{\label{sec:MR}Magnetostatic Resonance: Energy Transfer and
Dynamical Chaos}

In this Section we examine dynamics of an individual test electron
in the magnetostatic field~\eqref{HybPump_Field} given by
system~\eqref{eqn:ZO_Eqns_Motion}. In components the equations
have the form:
\begin{eqnarray}
\ddot{x}_{0} & - & \omega_{\|}\dot{y}_{0} =
{}-\omega_{\perp}\dot{z}_{0}\sin(\omega_{0}z_{0}/V_{\|}), \nonumber \\%
\ddot{y}_{0} & + & \omega_{\|}\dot{x}_{0} = 0, 
\label{eqn:ZO_Eqns_Motion_Comps} \\%
\ddot{z}_{0} & = & \omega_{\perp}\dot{x}_{0}
\sin(\omega_{0}z_{0}/V_{\|}), \nonumber
\end{eqnarray}
where $\omega_{0} = 2\pi V_{\|}/\lambda_{w}$ is the undulator frequency and
$\omega_{\perp ,\|} = |e|B_{\perp ,\|}/(m_{e}c\gamma_0)$;
the overdots denote differentiation with respect to the transit
time $\tilde{t}$. One can easily find two first integrals of this
nonlinear dynamical system just by integrating its first two
equations and take the energy conservation law as the third
linearly independent one
\begin{eqnarray}
\dot{x}_{0} - \omega_{\|}y_{0} -
(2\pi)^{-1}\omega_{\perp}\lambda_{w}\cos(\omega_{0}z_{0}/V_{\|})
& \equiv & F_{1} = \mbox{const}, \nonumber \\%
\dot{y}_{0} + \omega_{\|}x_{0} \equiv F_{2} = \mbox{const}, 
\,\,%
\dot{z}_{0}^{2} + \dot{x}_{0}^{2} + \dot{y}_{0}^{2} & \equiv &
F_{3} = \mbox{const}. \nonumber
\end{eqnarray}
Although, system~\eqref{eqn:ZO_Eqns_Motion_Comps} cannot be
integrated in quadratures everywhere because it possesses non-zero
Lyapunov exponents (see below Fig.~\ref{fig:LyapunovExponent}
and~\cite[p.~208]{LiebLikht1992}).
It should be noted that out of six quantities $\lambda_{w}$,
$B_{\perp}$, $B_{\|}$, $V_{\|}$, $V_{x}$ and $V_{y}$ only four
their dimensionless combinations $\varepsilon =
\omega_{\perp}/\omega_{0}\equiv c\mathcal{K}/(\gamma_{0}V_{\|})$
($\mathcal{K}$ is the conventional undulator parameter),
$\sigma_{0} = \omega_{\|}/\omega_{0}$, $V_{x}/V_{\|}$ and
$V_{y}/V_{\|}$ define the behavior of nonlinear dynamic
system~\eqref{eqn:ZO_Eqns_Motion_Comps}. Solution to
system~\eqref{eqn:ZO_Eqns_Motion_Comps} can be obtained by the use
of a method of Lindshtedt~\cite{Linshtedt1883,Blaquiere1966} of
asymptotic expansion of trajectory and frequencies in the small parameter
$\varepsilon$ (see~\cite{US2004} for a detailed exposition of the
method in the case $V_{x}=V_{y}=0$). To the order
$o(\varepsilon^{2})$ the velocity components read:
\begin{widetext}
\begin{eqnarray}\label{ZO_Eqns_Motion_Sols}
\dot{x}_{0}(\tilde{t}) & = & \hphantom{-}
V_{\perp}\cos(\Omega_{\|}\tilde{t} -\psi) +
\bar{v}_{\|}\frac{\omega_{\perp}\Omega_{0}}{\Omega_{0}^{2} -
\Omega_{\|}^{2}}\Bigl[\cos(\Omega_{0}\tilde{t}) -
\cos(\Omega_{\|}\tilde{t})\Bigr], \nonumber \\
\dot{y}_{0}(\tilde{t}) & = & - V_{\perp}\sin(\Omega_{\|}\tilde{t}
-\psi) + \bar{v}_{\|}\frac{\omega_\perp
\Omega_{0}}{\Omega_{0}^{2} - \Omega_{\|}^{2}}\Bigl[
\sin(\Omega_{\|}\tilde{t}) -
\frac{\Omega_{\|}}{\Omega_0}\sin(\Omega_{0}\tilde{t})\Bigr], \\
\dot{z}_{0}(\tilde{t}) & = & \bar{v}_{\|} -
V_{\perp}\omega_{\perp} \Bigl[\frac{\cos([\Omega_{0} -
\Omega_{\|}]\tilde{t} +\psi)}{2(\Omega_{0} - \Omega_{\|})} +
\frac{\cos([\Omega_{0} + \Omega_{\|}]\tilde{t} -
\psi)}{2(\Omega_{0} + \Omega_{\|})}\Bigr], \nonumber
\end{eqnarray}
\end{widetext}
where $V_{\perp}=(V_{x}^{2} + V_{y}^{2})^{1/2}$, $\sin{\psi} =
V_{y}/V_{\perp}$, $\bar{v}_{\|} = \kappa V_{\|}$, $\Omega_{0} =
\kappa\omega_{0}$ and $\Omega_{\|} =
\sigma\omega_{\|}/\sigma_{0}$. Here $\kappa$ and $\sigma$ to the
order $o(\varepsilon^3)$ are given by the following
self-consistent system of algebraic equations:
\begin{widetext}
\begin{equation}\label{Kappa_and_Phi}
\kappa = 1 + \frac{\varepsilon V_{x}}{V_{\|}(\kappa^2-\sigma^2)} -
\varepsilon^{2}\Bigr[\frac{3\kappa^2 + \sigma^2}{4(\kappa^2
-\sigma^2)^2} - \frac{V_{x}^{2}(8\kappa^4-13\kappa^2\sigma^2 +
\sigma^4)}{8V_{\|}^{2}\kappa^2(\kappa^2 -\sigma^2)^3} +
\frac{V_{y}^{2}(2\kappa^4 +
11\kappa^2\sigma^2-\sigma^4)}{8V_{\|}^{2}\kappa^2(\kappa^2
-\sigma^2)^3} \Bigl], \frac{\sigma}{\sigma_{0}} = 1 +
\frac{\varepsilon^2(\kappa^2 + \sigma^2)}{4(\kappa^2
-\sigma^2)^2};
\end{equation}
\end{widetext}
$\kappa$ and $\sigma$ are found as functions of $\varepsilon$ and
$\sigma_{0}$ with $V_{x}/V_{\|}$ and $V_{y}/V_{\|}$ as parameters.
An analytical solution to Eqs.~\eqref{Kappa_and_Phi} can be found
by successive iterations starting with 'non-renormalized' values
$1$ and $\sigma_{0}$ for $\kappa$ and $\sigma$, respectively. It
is worth noting that Eqs.~\eqref{Kappa_and_Phi} order by order in
$\varepsilon$ provide cancelation of secular terms in the solution
procedure for system~\eqref{eqn:ZO_Eqns_Motion_Comps}
(cf.~\cite{US2004}, which contains the limit of
\eqref{Kappa_and_Phi}, \eqref{ZO_Eqns_Motion_Sols} and
\eqref{ZO_Eqns_Motion_SolsAdds} for the case $V_{x}=V_{y}=0$). For
completeness, expressions for contributions to the velocities
\eqref{ZO_Eqns_Motion_Sols} of the order
$\mathcal{O}(\varepsilon^{2})$ are entered in the
Appendix\ref{App:Velocity}. Using
Eqs.~\eqref{Kappa_and_Phi},~\eqref{ZO_Eqns_Motion_Sols} and
\eqref{ZO_Eqns_Motion_SolsAdds} one can check that
$\vec{v}_{0}^{\,2}$ is conserved to the order $o(\varepsilon^{3})$
and is equal to $V_{x}^{2} + V_{y}^{2} + V_{\|}^{2}$ as required
by the conservative nature of system~\eqref{eqn:ZO_Eqns_Motion_Comps}.
It should be also noted that this aggregate solution to the order
$o(\varepsilon^{3})$ is valid not only for the `ideal' hybrid
planar magnetostatic field~\eqref{HybPump_Field} but
also for its 'realizable' counterpart involving
$\vec{B}_{w}(\vec{r})$. The trajectories of electrons are then
given by a straightforward integration.

From~\eqref{ZO_Eqns_Motion_Sols}
and~\eqref{ZO_Eqns_Motion_SolsAdds} it follows that the motion of
electron is a superposition of constant motion with velocity
$\bar{v}_{\|}$ and three-dimensional oscillations with normal
undulator $\Omega_{0}$ and cyclotron $\Omega_{\|}$ frequencies,
which differ from their partial analogues $\omega_{0}$ and
$\omega_{\|}$ by `renormalization' multipliers $\kappa$ and
$\sigma/\sigma_{0}$. Recall that each subsystem of a nonlinear
dynamical system possesses its own (partial) oscillation
frequency if interaction between subsystems vanishes. Non-zero
interaction modifies those oscillation frequencies by some
amount~\cite{RabTrub1989,Arnold1993} (represented here by
multipliers $\kappa$ and $\sigma/\sigma_{0}$); $\kappa$ also shows
what portion of electron initial kinetic energy is transferred to
the constant motion.

As $\Omega_{\|}$ tends to $\Omega_0$ the amplitudes of transversal
and longitudinal oscillations increase. This signifies the existence
of an internal resonance~\cite{ZaslSagd1988}, i.e. a resonant transfer
of kinetic energy from longitudinal constant motion to
transversal and longitudinal oscillatory degrees of freedom. We call
this situation magnetostatic resonance or \textit{magnetoresonance}
as is customary in the FEM theory~\cite{Marshall1983}.

\subsection{Zero initial transversal velocity}

\begin{figure}[t]
\scalebox{1.0}{\includegraphics[0,0][244,163]{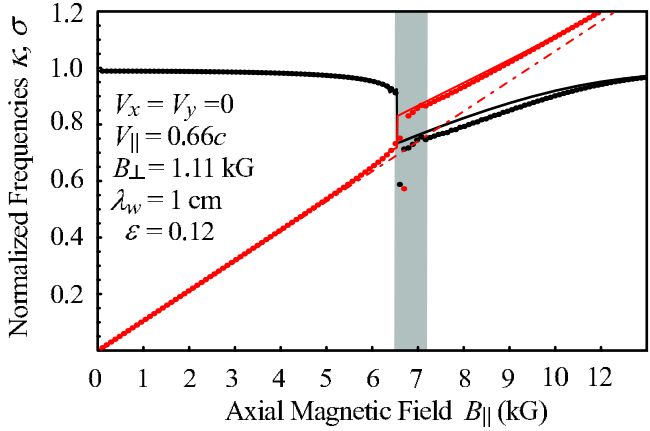}}
\caption{\label{fig:KappaSigma_Full} $\kappa$ and $\sigma$ as
functions of the guide magnetic field $B_{\|}$. $\kappa$ and and
$\sigma$ are depicted in black and red colors, respectively; for
reference $\sigma_{0}(B_{\|})$ is given by the chain line. The
darkened area corresponds to the zone of chaotic dynamics as given
by the Chirikov resonance-overlap
criterium~\eqref{eqn:ChirikovCriterium}.}
\end{figure}

\begin{figure}[t]
\scalebox{1.0}{\includegraphics[0,0][245,163]{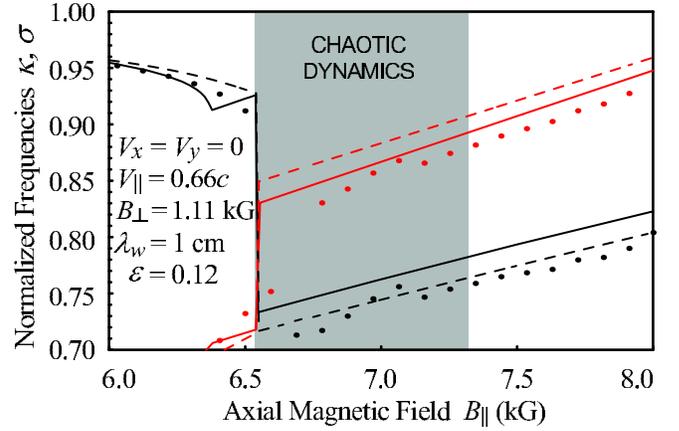}}
\caption{\label{fig:KappaSigma_MRZone} $\kappa$ and $\sigma$ as
functions of the guide magnetic field for $B_{\|}$ values around the
magnetoresonance. $\kappa$ and  and $\sigma$ are depicted in black
and red colors, respectively; for reference $\sigma_{0}(B_{\|})$ is
given by the chain line.}
\end{figure}

\begin{figure}[b]
\scalebox{1.0}{\includegraphics[0,0][237,290]{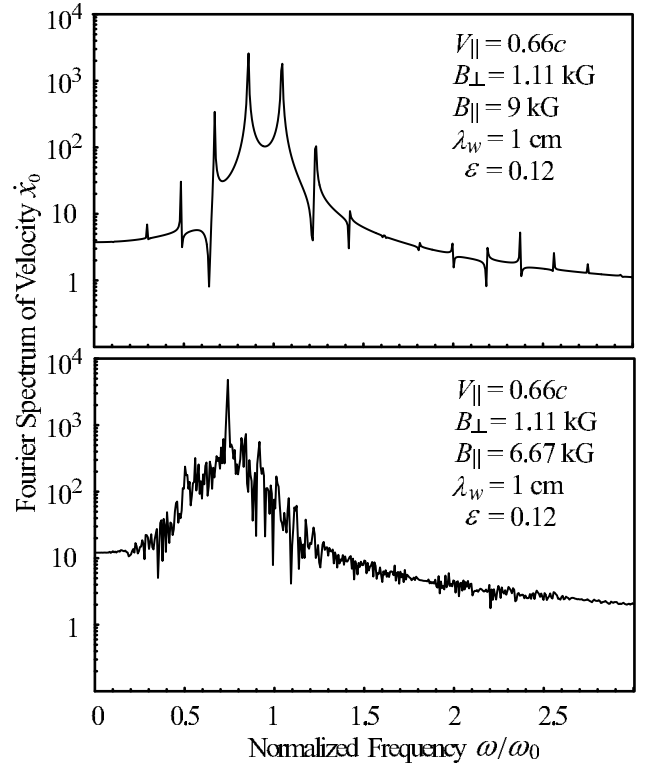}}
\caption{\label{fig:FourierSpectra} Logarithmically scaled Fourier
spectra of $\dot{x}_{0}$ as function of normalized frequency
$\omega/\omega_0$. Top figure shows regular dynamics while bottom
figure corresponds to chaotic one.}
\end{figure}

In the case when initial transversal velocity of electrons
vanishes ($V_{x} = V_{y} = 0$), the expressions
\eqref{ZO_Eqns_Motion_Sols}, \eqref{Kappa_and_Phi} and
\eqref{ZO_Eqns_Motion_SolsAdds} are greatly simplified. Then, as
mentioned above, the single-particle electron dynamics is defined
only by two dimensionless quantities $\varepsilon$ and
$\sigma_{0}$. To calculate the normal undulator and cyclotron
frequencies $\Omega_{0}$ and $\Omega_{\|}$, we need to solve for
$\kappa$ and $\sigma$ the system of two algebraic equations
\eqref{Kappa_and_Phi}. To achieve maximal accuracy it turns out
advantageous to solve this system numerically and use thus
obtained values of $\kappa$ and $\sigma$ in the analytical
calculations. In Fig.~\ref{fig:KappaSigma_Full} the results of
such a solution are shown in firm lines. For comparison, the
result of calculation of $\kappa$ and $\sigma$ via Fourier
analysis of direct numerical solution of
Eqs.~\eqref{eqn:ZO_Eqns_Motion_Comps} for the velocity components is
also given (shown by dots in Fig.~\ref{fig:KappaSigma_Full}). In
Fig.~\ref{fig:KappaSigma_MRZone} the graphs of $\kappa$ and
$\sigma$ vs. $B_{\|}$ are shown in more details for the guide
magnetic field values around the magnetoresonance. At the
magnetoresonance $B^{\mathrm{res}}_{\|} \approx 6.67$~kG the
normal frequencies $\Omega_{0}$ and $\Omega_{\|}$ experience jumps
in order to comply with the conservation of kinetic energy in the
dynamical system \eqref{eqn:ZO_Eqns_Motion_Comps}. As $\varepsilon$
grows above a certain threshold value two zones of regular
dynamics become separated by a chaotic zone centered about the
magnetoresonant value of the guide magnetic field (see
Fig.~\ref{fig:ContourPlot_ZTV}). Typical Fourier spectra of
velocity are shown in Fig.~\ref{fig:FourierSpectra}. Away from the
magnetoresonant values of the guide magnetic field one can easily
identify two distinct frequencies corresponding to $\Omega_{0}$
and $\Omega_{\|}$, their beating frequencies and harmonics. Around
the magnetoresonance the spectrum is nearly continuous, which is
one of the criteria of dynamical chaos.

Mutual orientation of the undulator and guide magnetic fields
causes the kinetic energy transfer from the longitudinal constant
motion to undulator vibrational (frequencies $\omega_{0}$ and
$\Omega_{0}$) and then to cyclotron vibrational (frequencies
$\omega_{\|}$ and $\Omega_{\|}$) degrees of freedom of the
nonlinear dynamical system \eqref{eqn:ZO_Eqns_Motion_Comps}; the
coupling is provided by the amplitude of undulator magnetic field
$B_{\perp}$ ($\varepsilon = \omega_{\perp}/\omega_{0}$).
Such a picture allows one two possible ways of energy transfer: at
magnetoresonant ($\Omega_{0}\approx \Omega_{\|}$) and
non-magnetoresonant ($\Omega_{0}\not\approx \Omega_{\|}$) coupling
regimes. Obviously, the magnetoresonant regime requires smaller
coupling and may become more advantageous for microwave FEM
amplifiers and oscillators. It is then necessary to clarify how
accessible are these regimes of FEM operation.

\begin{figure}[t]
\scalebox{1.0}{\includegraphics[0,0][241,160]{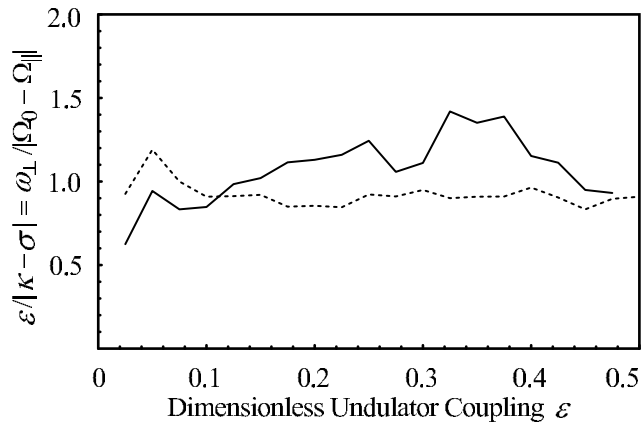}}
\caption{\label{fig:MaxValueMRM} Maximal value of magnetoresonant
multiplier as function of dimensionless coupling $\varepsilon$.}
\end{figure}

Viewing dynamical system~\eqref{eqn:ZO_Eqns_Motion_Comps} as a
nonlinear pendulum with external modulation of its vibration
amplitude (cf. the last equation
in~\eqref{eqn:ZO_Eqns_Motion_Comps}), one can estimate the
condition of appearing of chaotic layer near the separatrix. An
analytical estimate is consistent with numerical calculations and
coincides with the Chirikov resonance-overlap
criterium~\cite{Chirikov1979}:
\begin{equation}\label{eqn:ChirikovCriterium}
|\Omega_{0} - \Omega_{\|}| < \omega_{\perp}\,\, (|\kappa - \sigma| <
\varepsilon),
\end{equation}
i.e. appearance of the chaotic state takes place whenever the
difference between normal frequencies of the system becomes less
than the coupling. As will be shown in Section~\ref{sec:LTMA}, the
absolute value of the ratio $\omega_{\perp}/(\Omega_{0} -
\Omega_{\|})$ (called below the \textit{magnetoresonant}
\textit{multiplier}) defines the microwave gain. Results of
numerical calculations of its maximal value for zones of regular
dynamics and practically accessible $\varepsilon(B_{\perp},V_{\|})$
are given in Fig.~\ref{fig:MaxValueMRM} (for each value one needs to
optimize $B_{\|}$; solid and dotted lines correspond to approach of
the chaotic zone from low and high values of the guide magnetic
field, respectively). This shows that in the magnetoresonant case
there exists limitation on the fraction of energy transferred from
constant motion to oscillatory (transversal and longitudinal)
degrees of freedom posing a fundamental limit on an FEM efficiency.
A calculation shows that $\overline{\dot{x}_{0}^{2}(\tilde{t}) +
\dot{y}_{0}^{2}(\tilde{t}\,)}/\overline{\dot{z}_{0}^{2}(\tilde{t}\,)}
\approx 0.48$ (bars denote time averaging). In
Fig.~\ref{fig:ContourPlot_ZTV} we present a contour plot of
$\kappa\equiv \bar{v}_{\|}/V_{\|}$ as function of dimensionless
coupling $\varepsilon$ and guide magnetic field $\sigma_{0}$. The
region of chaotic dynamics is given in black. In
Fig.~\ref{fig:LyapunovExponent} a  map of the major Lyapunov
exponent as function of $\varepsilon$ and $\sigma_{0}$ is shown. The
calculation included a check that the sum of all Lyapunov exponents
is equal to zero as mandatory for a conservative dynamical system
because of the Liouville theorem. %

\subsection{Non-zero initial transversal
            velocity;\\ chaos control}\label{subsec:NZ_ITV}

Let us start with demonstration of some features of
solutions~\eqref{ZO_Eqns_Motion_Sols} for a non-zero initial
transversal velocity. As follows from~\eqref{ZO_Eqns_Motion_Sols}
and~\eqref{Kappa_and_Phi}, one can have either suppression ($V_{x}
> 0$) or enhancement ($V_{x} < 0$) of the velocity amplitude of
cyclotron vibrations, respectively. In particular, for $V_{y} = 0$
and $V_{x} = \bar{v}_{\|}\omega_{\perp}\Omega_{0} /(\Omega_{0}^{2}
- \Omega_{\|}^{2})$ there occurs a \textit{complete}, i.e. to the
first order in the small parameter $\varepsilon$, suppression of
cyclotron vibrations in the transverse to the $z$-axis plane. In
Fig.~\ref{fig:VelAmpl_Sup/Enh} this situation is demonstrated for
$V_{y} = 0$ and both signs of $V_{x}$ of the same absolute value
$|V_{x}| = 0.1c$, $V_{\|} = 0.66 c$, $B_{\perp} = 1.11$~kG, the
value of $B_{\|}$ is chosen to be $6.67$~kG and corresponds to the
magnetoresonance for $V_{x} < 0$ (only projections of electron
trajectories in the $xy$-plane are shown).
\begin{figure}[b]
{\includegraphics[233,133][471,366]{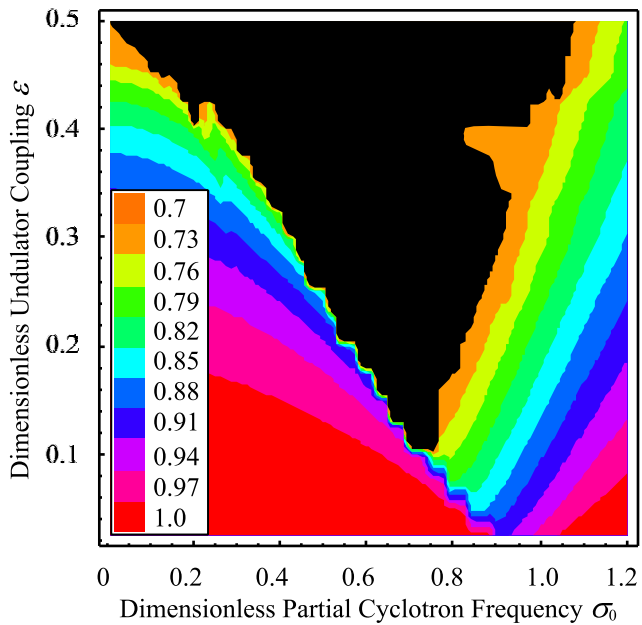}}
\caption{\label{fig:ContourPlot_ZTV} Contour plot of $\kappa\equiv
\bar{v}_{\|}/V_{\|}$ as function of dimensionless coupling
$\varepsilon$ and guide magnetic field $\sigma_{0}$. The region of
chaotic dynamics is given in black.}
\end{figure}

It should be noted that in general the presence of non-zero initial
transversal velocity influences the shape of chaotic dynamics zone.
The influence of $V_y$ is symmetric regarding its sign and, hence,
less relevant while that of $V_x$ is much more important (see
\eqref{Kappa_and_Phi}). In particular, the $x$-component of initial
velocity of a certain sign can either suppress ($V_x > 0$) or
enhance ($V_x < 0$) the zone of chaotic dynamics as shown in
Figs.~\ref{fig:ContourPlot_Sup} and~\ref{fig:ContourPlot_Enh}.
Non-zero $V_x$ also changes position of the edge of regular dynamics
zone at zero guide magnetic field. This is due to the fact that
under such conditions position of separatrix on the phase plane
$z_{0}$-$\dot{z}_{0}$ depends on the value and sign of $V_x$ and is
independent of $V_y$. Position of the edge at zero guide magnetic
field ($\sigma_{0} = 0$) is calculated to be $\varepsilon = ([1 +
(V_{x}/V_{\|})^2]^{1/2} + V_{x}/V_{\|})/2$.

Having considered separately the particular cases of zero and
non-zero initial transversal velocity, below we will study some
generic properties of nonlinear dynamical system
\eqref{eqn:ZO_Eqns_Motion_Comps}. First, it should be emphasized
that nonlinear dynamical system \eqref{eqn:ZO_Eqns_Motion_Comps} is
degenerate in the sense that at zero amplitude of undulator magnetic
field or zero guide magnetic field the dynamics is characterized by
only one normal frequency instead of two. Under such conditions
(after subtraction of constant motion along the $z$-axis), electron
trajectories constitute invariant curves, while in the generic case
they wind up (surfaces diffeomorphic to) two-dimensional invariant
tori (see below). Second, an important feature of solutions in the
cases of zero and non-zero initial transversal velocity consists in
cancelation of non-physical secular terms in the series expressions.
This is due to the use of a method of Lindshtedt for finding
electron trajectories. Third, as mentioned at the beginning of this
Section, nonlinear dynamical system \eqref{eqn:ZO_Eqns_Motion_Comps}
possesses three linearly independent first integrals $F_i$, $(i = 1,
2, 3)$ that are not all in involution
\begin{equation}\label{eqn:PoissonBrackets}
\{F_{1},F_{2}\} = -\omega_{\|},\quad
\{F_{1},F_{3}\}=\{F_{2},F_{3}\}=0.
\end{equation}
($\{F_{i},F_{j}\}$ denote the Poisson brackets). Therefore, it does
not fall under conditions of the Liouville theorem. However,
observing that the identity, $I$, is also a necessary member\hfill
of the Lie\hfill algebra\hfill of first\hfill integrals of\hfill system %
\begin{figure}[h!]
{\includegraphics[230,103][469,336]{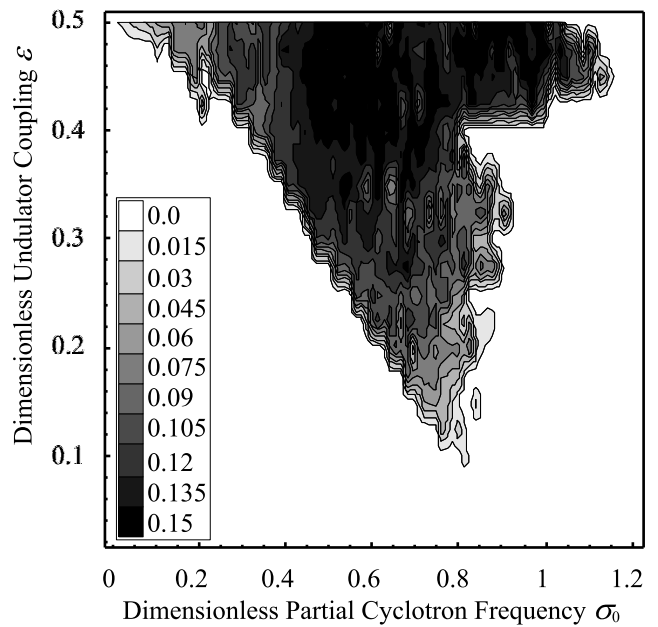}}
\caption{\label{fig:LyapunovExponent} Map of the major Lyapunov
exponent.}
\end{figure}%
{}\\%
{}\\%
\eqref{eqn:ZO_Eqns_Motion_Comps} by virtue of the first relation in
\eqref{eqn:PoissonBrackets}, one can infer that the condition of
non-commutative integrability holds for the system under
investigation (for details see~\cite[p.~190]{Kozlov1998}). It then
follows that we can apply the Nekhoroshev theorem to find the
dimensions of invariant tori (the loci of the electrons'
trajectories), which foliate the system's phase space,
\cite{Nekhoroshev1972}. %
\begin{figure}[t]
\includegraphics[0,0][264,192]{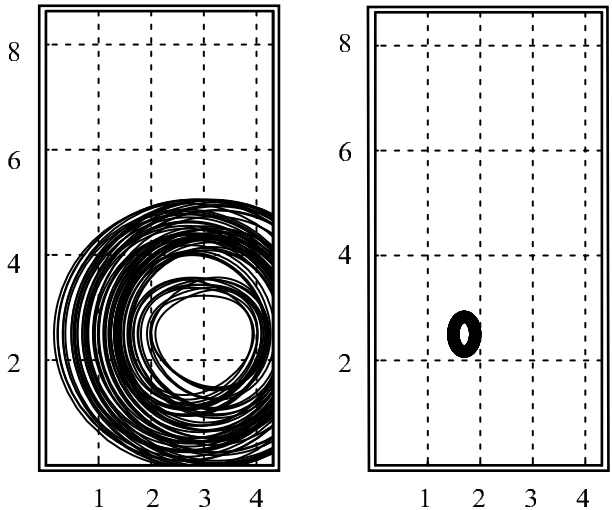}
\caption{\label{fig:VelAmpl_Sup/Enh} Projections of electron
trajectories in $xy$-plane. (a) $V_{x} = -0.1c$; (b) $V_{x} = 0.1c$.
Bounding box is of a standard waveguide size $a\times b$ = $8.6$~mm
$\times$ $4.3$~mm.}
\end{figure}%
\begin{figure}[b]
\includegraphics[216,124][454,357]{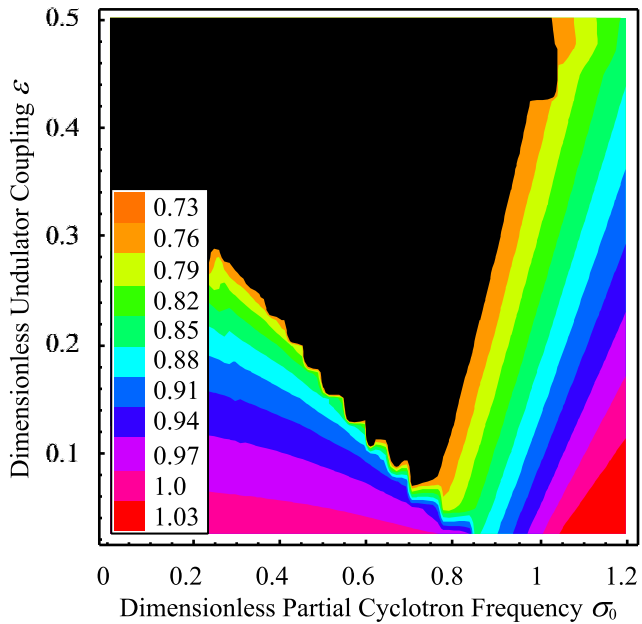}
\caption{\label{fig:ContourPlot_Sup} Contour plot of $\kappa\equiv
\bar{v}_{\|}/V_{\|}$ as function of $\varepsilon$ and $\sigma_{0}$:
suppression of zone of chaotic dynamics ($V_{x}=0.3V_{\|}$). The
region of chaotic dynamics is given in black.}
\end{figure}%
There exist four first integrals ($I$, $F_{i}, i=1,2,3$) for a three
dimensional nonlinear dynamical
system~\eqref{eqn:ZO_Eqns_Motion_Comps}, two of which ($I$ and
$F_{3}$) constitute a complete set of independent first integrals in
involution. Then, the dimensions of invariant tori are equal to two.
Under natural projection into the system's configuration space one
can visualize such two-dimensional tori. Namely, in the reference
frame, which moves with the mean velocity $\bar{v}_{\|}$ of electron
(see Eqs.~\eqref{ZO_Eqns_Motion_Sols}
and~\eqref{ZO_Eqns_Motion_SolsAdds}), $\bar{\gamma}_{\|} = (1 -
\bar{v}^{2}_{\|}/c^{2})^{-1/2}$, its trajectory winds up a (surface
diffeomorphic to) two-dimensional torus
(Fig.~\ref{fig:TorusRegular}). Such a regular behavior is exhibited
by the dynamical system under investigation only for particular
range of dimensionless parameters $\varepsilon$, $\sigma_{0}$,
$V_{x}/V_{\|}$ and $V_{y}/V_{\|}$ (cf.~\cite{Kozlov1983}) as can be
understood from
Figs.~\ref{fig:ContourPlot_ZTV},~\ref{fig:ContourPlot_Sup}
and~\ref{fig:ContourPlot_Enh}. When parameters change in such a way
that the motion of electrons becomes chaotic, we can observe the
disintegration of the corresponding invariant torus shown in
Figs.~\ref{fig:TorusDisintergSmallT}
and~\ref{fig:TorusDisintergLongT}. In
Fig.~\ref{fig:TorusDisintergSmallT} the trajectory of electron is
drawn for a small period of time and looks like winding up a
two-dimensional torus, while for a longer period of time it does not
(Fig.~\ref{fig:TorusDisintergLongT}).
\begin{figure}[b]
\includegraphics[216,124][454,357]{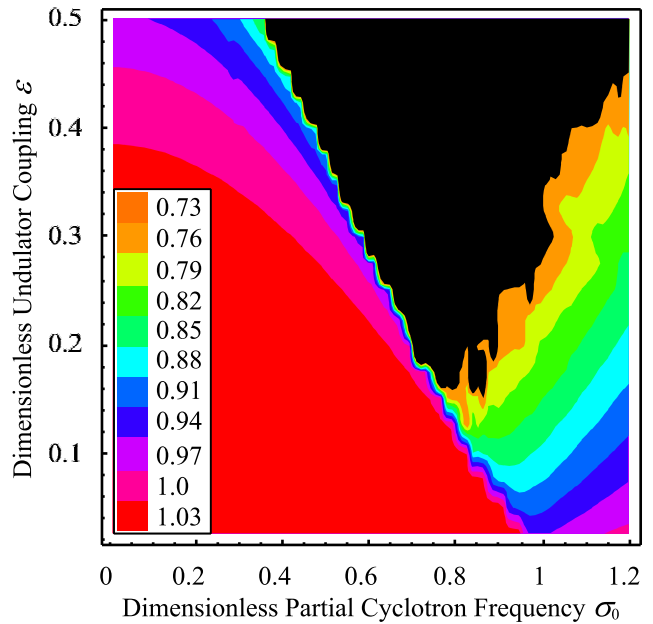}
\caption{\label{fig:ContourPlot_Enh} Contour plot of $\kappa\equiv
\bar{v}_{\|}/V_{\|}$ as function of $\varepsilon$ and $\sigma_{0}$:
enhancement of zone of chaotic dynamics  ($V_{x}={}-0.3V_{\|}$). The
region of chaotic dynamics is given in black.}
\end{figure}

Summarizing this Section, we observe that there are two regimes for
pumping kinetic energy into transversal degrees of freedom of a
hybrid planar FEM: the first one (advocated here as more, if not the
only, efficient one for the terahertz waveband) consists in low
amplitude of undulator magnetic field and the guide magnetic field
as close to its magnetoresonant value as only consistent with the
regular dynamics; the second regime (usually considered in the
literature) would also provide the same ratio
$\omega_{\perp}/(\Omega_{0} - \Omega_{\|})$ through a %
\begin{figure}
\includegraphics[0,0][172,286]{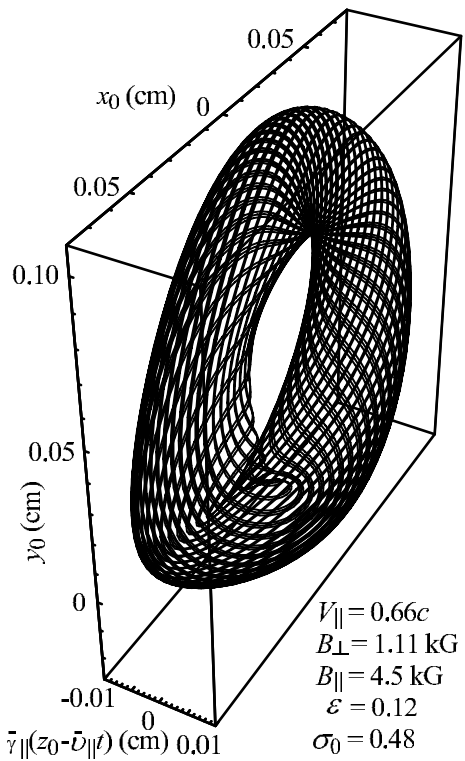}
\caption{\label{fig:TorusRegular} Electron trajectory (winding of
two-dimensional torus) under conditions of regular dynamics.}
\end{figure}%
\begin{figure}
\includegraphics[0,0][170,282]{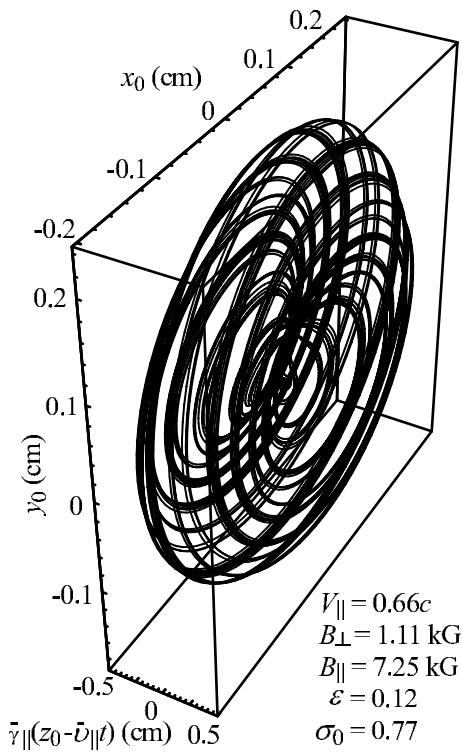}
\caption{\label{fig:TorusDisintergSmallT} Disintegration of
invariant torus for a short period of time.}
\end{figure}%
high absolute value of $|\Omega_{0} - \Omega_{\|}|$ given by an
off-resonant guide magnetic field $B_{\|}$ and a necessarily high
value of amplitude of undulator magnetic field $B_{\perp}$. It,
however, seems that values of $B_{\perp}$ in excess of $10$~kG are
currently practically unattainable while guide magnetic fields of up
to $50$~kG can be presently achieved.

\section{\label{sec:LTMA} Linear Amplification: Magnetostatic
         Resonance and Maximal Gain}

As mentioned in Sec.~\ref{sec:GE}, analytical knowledge of
electrons' trajectories~\eqref{ZO_Eqns_Motion_Sols} in %
\begin{figure}[b]
\includegraphics[0,0][171,283]{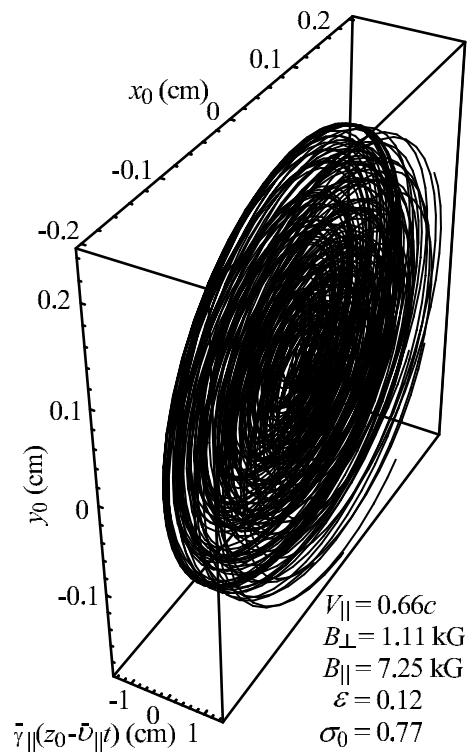}
\caption{\label{fig:TorusDisintergLongT} Disintegration of invariant
torus for a long period of time.}
\end{figure}%
magnetostatic field~\eqref{HybPump_Field} allows one to develop a
gyrotron-like analytical linear theory of microwave amplification.
In particular, in Ref.~\cite{RadioPhys2006} such an approach was
constructed for interaction of an electron beam with a mode of an
arbitrary regular waveguide in a hybrid FEM. It should be noted that
analytical theory of this type is valid not only far away from the
magnetoresonance ($\Omega_{0}\approx\Omega_{\|}$,
cf.~\cite{BratmanEtAl1978}) but also in its close vicinity. Here we
will apply this theory to study amplification of microwaves in a
hybrid planar FEM under conditions of magnetoresonance using as an
example amplification of $\mathrm{TE}_{10}$~and
$\mathrm{TE}_{01}$~modes of rectangular waveguide with wide $a$ and
narrow $b$ sides. Exactly these modes allow one to utilize the
advantage of enhanced current of sheet electron beams (cf., for
example~\eqref{eqn:GrowthRate}). Specifically, in each case we need
to solve system~\eqref{FO_System} of ordinary differential equation
with quasi-periodic coefficients. Its exact solution is achieved by
application of a quasi-periodic analogue of Floquet's theorem and,
under some additional conditions, comes in the form of infinite
series in undulator and cyclotron
harmonics,~\cite[pp.~90-96]{Haken1983}. Under particular synchronism
conditions the series may be truncated to retain solution pertinent
to this very synchronism neglecting other combination ones (see
\cite{Gaponov1961,VanSoln1973}). In this case, we can write $C(z) =
C_{0} \exp{(i\delta k_{z}z)}$ for the microwave amplitude ($|\delta
k_{z}| \ll |k^{0}_z|$), where $\delta k_z$ is the spatial growth
rate.

\subsection{Undulator synchronism}\label{subsec:LTMA_US}

Let the undulator synchronism condition, \cite{US2006}, is
fulfilled
\begin{equation}\label{eqn:UndulatorSynch}
\omega - k_z^{0} \bar{v}_{\|}\approx\Omega_{0}.
\end{equation}
Using the fundamental $\mathrm{TE}_{10}$~mode ($k^{0}_{z} =
[(\omega/c)^2-(\pi/a)^2]^{1/2}$) of a rectangular waveguide as an
explicit demonstration example, we solve~\eqref{FO_Eqns_Motion} to
the order $o(\varepsilon^{2})$ (retaining only summands up to the
second order in $V_{\perp}/\bar{v}_{\|}$) and obtain
\begin{equation}\label{eqn:MicrowaveTrajectories}
\vec{r}_{1}(\tilde{t}) = r_1\vec{e}_r + %
r_0\phi_1\vec{e}_\phi + z_1\vec{e}_z,
\end{equation}
where
\begin{widetext}
\begin{eqnarray*}
r_1 & = & -\frac{i\pi\lambda_{w}|e|C_{0}}{4m_e c\,\gamma_{0}}
    \frac{\omega_{\perp}\Omega_{0}}{\Omega^{2}_{0} - \Omega^{2}_{\|}}
    \frac{\exp(i[\delta k_z\bar v_{\|} -
    \delta\omega]\tilde{t})}{\delta k_z\bar v_{\|} - \delta\omega}
    \Bigl\{\sin(\Omega_{\|}\tilde{t})\cos\Bigl(\frac{\pi}{a}\bigl[\bar{x} +
    \frac{V_y}{\Omega_{\|}}\bigr]\Bigr) + \Bigr.\\
& & \hspace{3.6cm}
\Bigl.\frac{ia\lambda_{w}\Omega^{2}_{\|}\Omega_{0}V_{\perp}}{2\pi^{2}c^2(\Omega_{0}
    - \Omega_{\|})\bar{v}_{\|}}\sin\Bigl(\frac{\pi}{a}\bigl[\bar{x} +
    \frac{V_y}{\Omega_{\|}}\bigr]\Bigr) - \frac{\lambda_{w}^2\Omega_0^2
    V_\perp^2}{16a^2\Omega_{\|}^2\bar{v}_{\|}^2}\cos(\Omega_{\|}\tilde{t} -
    \psi)\cos\Bigl(\frac{\pi}{a}\bigl[\bar{x} +
    \frac{V_y}{\Omega_{\|}}\bigr]\Bigr)\Bigr\}, \\
r_0\phi_{1} & = & -\frac{i\pi\lambda_{w}|e|C_{0}}{4m_e
    c\,\gamma_{0}}\frac{\omega_{\perp}\Omega_{0}}{\Omega^{2}_{0}
    - \Omega^{2}_{\|}}\frac{\exp(i[\delta k_z\bar v_{\|} -
    \delta\omega]\tilde{t})}{\delta k_z\bar v_{\|} - \delta\omega}\Bigl\{
    \cos(\Omega_{\|}\tilde{t})\cos\Bigl(\frac{\pi}{a}\bigl[\bar{x} +
    \frac{V_y}{\Omega_{\|}}\bigr]\Bigr) + \\
& & \hspace{3.1cm}\frac{a\lambda_{w}\Omega_{0}\Omega_{\|}
    (\omega + \Omega_{\|})V_{\perp}}{2\pi^{2}c^2
    (\delta k_z\bar v_{\|} - \delta\omega)\bar{v}_{\|}}
    \sin\Bigl(\frac{\pi}{a}\bigl[\bar{x} +
    \frac{V_y}{\Omega_{\|}}\bigr]\Bigr) + \frac{\lambda_{w}^2\Omega_0^2
    V_\perp^2}{16a^2\Omega_{\|}^2\bar{v}_{\|}^2}\sin(\Omega_{\|}\tilde{t} -
    \psi)\cos\Bigl(\frac{\pi}{a}\bigl[\bar{x} +
    \frac{V_y}{\Omega_{\|}}\bigr]\Bigr)\Bigr\}, \\
z_{1} & = & \frac{i\pi a|e|\omega C_{0}}{2m_e
    c\,\gamma_{0}^{3}}\frac{\omega_{\perp}\Omega_{\|}}{\Omega^{2}_{0}
    - \Omega^{2}_{\|}}\frac{\exp(i[\delta k_z\bar v_{\|} -
    \delta\omega]\tilde{t})}{(\delta k_z\bar v_{\|} -
    \delta\omega)^{2}}\Bigl\{1 - \frac{\lambda_{w}^2\Omega_0^2(\omega
    + \Omega_{\|})V_\perp^2}{16a^2\Omega_{\|}^2\bar{\gamma}^2_{\|}
    \omega\bar{v}_{\|}^2}\Bigr\}\sin\Bigl(\frac{\pi}{a}\bigl[\bar{x}
    + \frac{V_y}{\Omega_{\|}}\bigr]\Bigr),
\end{eqnarray*}
\end{widetext}
and $\delta\omega = \omega - k_z^0\bar{v}_\parallel - \Omega_0$ is
a small mismatch from the ideal synchronism ($\delta\omega = 0$).
It then follows that a major contribution to the spatial growth
rate of the waveguide mode exists even for an electron beam with
zero initial transverse velocity. For conciseness we will assume
$V_{\perp} = 0$ down to the end of this subsection. On the one
hand for $V_{\perp} = 0$
expressions~\eqref{eqn:MicrowaveTrajectories} describe a
well-known result (see~\cite{Freund1985}) that interaction between
the transverse component of oscillatory motion and microwave
results primarily in axial bunching, which becomes the main source
of instability (if $V_{\perp}$ equals to zero then radial and
azimuthal bunching is weaker because in this case $r_{1}$ and
$r_{0}\phi_{1}$ are inversely proportional to the small quantity
$(\delta k_z\bar{v}_{\|} - \delta\omega)$ while $z_1$ is inversely
proportional to its square). On the other hand (unlike
in~\cite{Freund1985}) we infer this from \textit{analytical}
expressions~\eqref{eqn:MicrowaveTrajectories}, which are valid for
all regular dynamics parameters as found in Sec.~\ref{sec:MR} and,
specifically, near the magnetoresonance.
Substituting~\eqref{eqn:MicrowaveTrajectories}
in~\eqref{FO_Eqns_Excitation}, one then gets the dispersion
equation for the $\mathrm{TE}_{10}$~mode.

Results of the outlined calculation in the case $V_{x} = V_{y} =
0$ are as follows:
\begin{widetext}
\begin{equation}\label{eqn:DispersionUndul}
\delta k_z (\delta k_z - \frac{\delta \omega}{\bar{v}_{\|}})^2 =
\mp \frac{\pi |I_0|}{2c\bar{U}\gamma_{0}^{3}}
\frac{\omega^2_\perp}{(\Omega_0^2 - \Omega_{\|}^2)^2}\times\left[
\begin{array}{r@{,\quad}l@{\;}c@{;}}
{\displaystyle \frac{\Omega_{\|}^2\omega}{abc}
\sin^{2}\Bigl(\frac{\pi\bar{x}}{a}\Bigr)} &
\mathrm{TE}_{10}\;\mbox{mode} & (k^{0}_{z} =
[(\omega/c)^2-(\pi/a)^2]^{1/2}) \\
{\displaystyle
\frac{\Omega_0^2\omega}{abc}\sin^{2}\Bigl(\frac{\pi\bar{y}}{b}\Bigr)}
& \mathrm{TE}_{01}\;\mbox{mode} & (k^{0}_{z} =
[(\omega/c)^2-(\pi/b)^2]^{1/2})
\end{array}
\right.
\end{equation}
\end{widetext}
where $\bar{U} = m_e\bar{v}_{\|}^2/(2 |e|)$ is the
non-re\-la\-ti\-vis\-tic beam voltage of constant motion. The
conditions for existence of two complex conjugated roots of each
of Eqs.~\eqref{eqn:DispersionUndul} consist in positivity of their
respective cubic discriminants. These yield in each case the
thresholds for such instabilities to occur,~\cite{IVEC2007}.

For the $\mathrm{TE}_{10}$~mode such a threshold condition reads
\begin{equation*}
\frac{\delta\omega}{\bar{v}_{\|}} > {}-\frac{3}{2
\gamma_0}\Bigl(\frac{\pi\omega
|I_0|}{abc^2\bar{U}}\Bigr)^{1/3}\Bigl(\frac{\omega_\perp
\Omega_\parallel}{\Omega_0^2 - \Omega_\parallel^2}\Bigr)^{2/3}
\sin^{2/3}\Bigl(\frac{\pi\bar{x}}{a}\Bigr).
\end{equation*}
For the ideal synchronism ($\delta\omega = 0$) the spatial growth rate of the
$\mathrm{TE}_{10}$~mode is given by
\begin{equation}\label{eqn:GrowthRate}
\mathrm{Im}\,\delta k_z\! =\!
\frac{\sqrt{3}}{2\gamma_0}\Bigl(\frac{\pi \omega
|I_0|}{2abc^2\bar{U}}\Bigr)^{\!1/3}\!\!
\Bigl(\frac{\omega_\perp\Omega_\parallel}{\Omega_0^2
-\Omega_\parallel^2}\Bigr)^{\!\!2/3}\!\!\!\sin^{2/3}\!
\Bigl(\frac{\pi\bar{x}}{a}\Bigr),
\end{equation}
$\mathrm{Re}\,\delta k_z$ provides a correction to the cold
propagation constant, $k^{0}_{z}$, caused by the presence of
electron beam. Result~\eqref{eqn:GrowthRate} is similar to those
found previously (cf., e.g.~\cite{Freund1986}) and also implies
that there could exist a substantial enhancement in the (microwave)
gain, $G = 8.63\,\mathrm{Im}\,\delta k_z$, because of the presence
of guide magnetic field. It is also significant that
\textit{analytical} expression~\eqref{eqn:GrowthRate} provides
a close approximation of the growth rate, which is in a good
agreement with the direct numerical simulations of the nonlinear
self-consistent system of relativistic equations of motion and
equations of excitation~\eqref{Master_Eqns}. This is mainly due to
the fact that in $\Omega_0$ and $\Omega_{\|}$ we account for the initial
electron velocity and magnitudes of magnetostatic fields not
only through the definitions of $\omega_0$ and $\omega_{\|}$ but
also via `renormalization' multipliers $\kappa$ and $\sigma$
(cf.~\eqref{Kappa_and_Phi} and Fig.~\ref{fig:KappaSigma_Full}).
More importantly, utilizing results obtained in Sec.~\ref{sec:MR},
we can provide an analytical estimate for the maximal attainable
gain. To accomplish this task let us substitute
from~\eqref{eqn:ChirikovCriterium} to find
\begin{equation}\label{eqn:GainOptimization}
\frac{\omega_{\perp}\Omega_{\|}}{\Omega^2_{0} - \Omega^2_{\|}} <
\frac{\Omega_{\|}}{\Omega_{0} + \Omega_{\|}}.
\end{equation}
Recall that there exists two ways of having the value of
magnetoresonant multiplier, $\omega_{\perp}/(\Omega_{0} -
\Omega_{\|})$, close to $1$: one
should either work far from the magnetoresonance (and provide for
a strong undulator amplitude $B_\perp$) or, as advocated at the
end of Sec.~\ref{sec:MR}, operate near the magnetoresonance
$\Omega_{0}\approx \Omega_{\|}$ with a moderate undulator
amplitude (see also Fig.~\ref{fig:MaxValueMRM}). From the
inspection of the right hand side of~\eqref{eqn:GainOptimization},
we observe that its value is about $1/2$ around the
magnetoresonance. This provides another reason for preferring a
hybrid planar FEM operation regime slightly above the
magnetoresonance in order to ensure the regular dynamics of
electron motion as detailed in Sec.~\ref{sec:MR} (see also
Figs.~\ref{fig:KappaSigma_Full},~\ref{fig:KappaSigma_MRZone}
and~\ref{fig:ContourPlot_ZTV}). We, therefore,
find for the upper limit on the maximal (near)~magnetoresonant gain of a
hybrid planar FEM under conditions of ideal undulator synchronism
($\delta\omega = 0$) and $V_\perp = 0$
\begin{equation}\label{magnetoresonat_growth rate}
G_{\mathrm{max}}^{\mathrm{res}} \approx
\frac{3.7}{\gamma_0}\Bigl(\frac{\pi \omega
|I_0|}{abc^2\bar{U}}\Bigr)^{1/3}\!\sin^{2/3}\Bigl(\frac{\pi\bar{x}}{a}\Bigr).
\end{equation}%
\begin{figure}[b!]
\includegraphics[57,441][301,583]{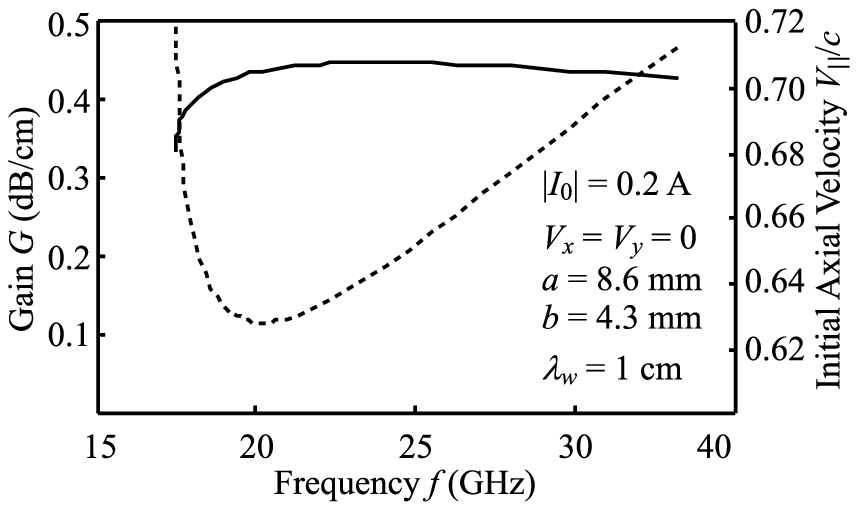}
\includegraphics[58,442][301,583]{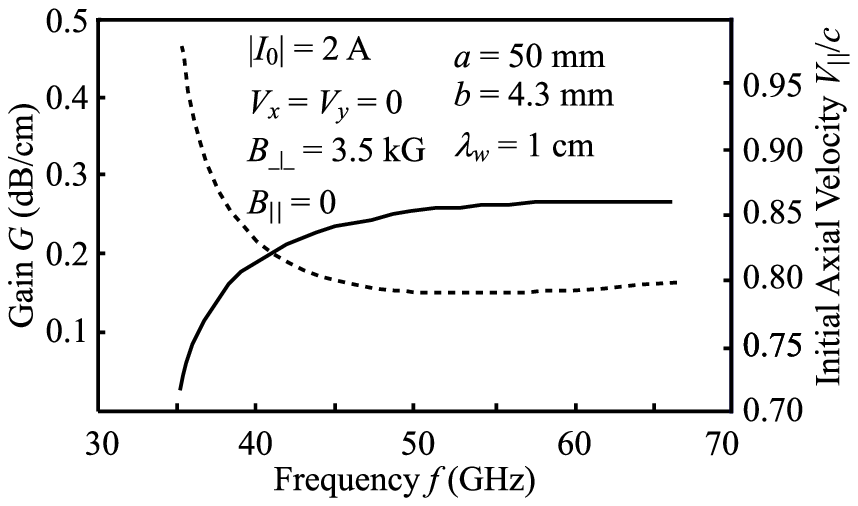}
\caption{\label{Broadband} In the top figure results of analytical
calculations of the maximal resonant gain (solid line) and initial
axial velocity of the beam (dashed line) for the fundamental
$\mathrm{TE}_{10}$~mode are presented. In the bottom figure we show
the gain (solid line) and initial axial beam velocity (dashed line)
at $B_{\|} = 0$ for the $\mathrm{TE}_{01}$~mode. All quantities are
shown as functions of amplified frequency.}
\end{figure}%
The same conditions of regular dynamics lead to an important
result that around the magnetoresonance the maximal gain is
independent of $B_\perp$ (although entering $\bar{v}_{\|}$ in
$\bar{U}$ through $\kappa$, it cancels out completely by virtue of
synchronism condition ($\bar{v}_{\|} = \omega/(k_z^0(\omega) +
2\pi/\lambda_{w})$); as a function of frequency,
$G_{\mathrm{max}}^{\mathrm{res}} \sim (k^{0}_z + 2\pi
/\lambda_{w})^{2/3}\bar{\gamma}^{-1}_{\|}\omega^{-1/3}$.

Under the assumption of undulator synchronism in the limit $B_{\|}
= 0$ ($\sigma = 0$ and $\kappa \approx 1 - 3\varepsilon^2/4$) an
electron beam without initial transverse velocity does not
interact with the $\mathrm{TE}_{m0}$~modes of a rectangular
waveguide. Then the $\mathrm{TE}_{01}$~mode turns out to be the
lowest one amplified by such an electron beam, and the bottom line
of Eqs.~\eqref{eqn:DispersionUndul} can be used to estimate
analytically the gain of a planar FEM~amplifier without the guide
magnetic field (cf.~\cite{Destler1996})
\begin{equation}\label{TE01_growth rate}
G|_{B_{\|}=0} = \frac{5.93}{\gamma_0} \Bigl(\frac{\pi\omega
|I_0|}{abc^2\bar{U}}\Bigr)^{1/3}
\Bigl(\frac{\omega_{\perp}}{\Omega_{0}}\Bigr)^{2/3}
\!\sin^{2/3}\Bigl(\frac{\pi\bar{y}}{b}\Bigr).
\end{equation}
Although, it should be noted that magnetoresonant gain ($\Omega_0
\approx \Omega_{\|}$ and $\omega_\perp/(\Omega_0 - \Omega_{\|})
\approx 1$) calculated using the bottom line of
Eqs.~\eqref{eqn:DispersionUndul} is always greater than that one
given by Eq.~\eqref{TE01_growth rate}.

Another important characteristic of an amplifier is its tunability.
In the case of undulator synchronism the frequency tuning of
an FEM~amplifier is achieved by changing initial axial velocity,
$V_\parallel$, of the electron beam and turns out to be limited
only by the requirement of single-mode operation regime. In
Fig.~\ref{Broadband} for the fundamental $\mathrm{TE}_{10}$~mode
one can see that the maximal resonant gain depends weakly on the
frequency in the operating range. To provide the single-mode
operation regime and interaction only with the forward wave, the
frequencies are chosen to range from $18$~GHz to $32$~GHz in this
case. Results of gain calculations for interaction with the
$\mathrm{TE}_{01}$~mode are also shown in Fig.~\ref{Broadband}.
The initial axial velocity must also change with the frequency
($\bar{v}_{\|} = \omega/(k_z^0(\omega) + 2\pi/\lambda_{w})$) as shown in
the both parts of Fig.~\ref{Broadband} to maintain the ideal
undulator synchronism (in the top figure for calculation of
$V_{\|}$ from a given $\bar{v}_{\|}$ we choose $\kappa = 0.8$, see
Fig.~\ref{fig:ContourPlot_ZTV}; in the bottom figure to obtain
$V_{\|}$ from a given $\bar{v}_{\|}$ one needs to solve the
equation $\bar{v}_{\|} = \kappa (V_{\|})V_{\|}$,
see~\eqref{Kappa_and_Phi} and note that in this case $\sigma = 0$,
$V_x = V_y = 0$).

\subsection{Hybrid synchronism}\label{subsec:LTMA_HS}

Under the condition of hybrid (undulator-cyclotron) synchronism,
\cite{US2006},
\begin{equation}\label{eqn:HybridSynch}
\omega - k_z^{0} \bar{v}_{\|} \approx \Omega_{0} + \Omega_{\|},
\end{equation}
we again can solve equation~\eqref{FO_Eqns_Motion} to the order
$o(\varepsilon^{2})$ retaining only summands up to the second
order in $V_{\perp}/\bar{v}_{\|}$ and those which contain the
square of the small quantity $(\delta k_z\bar{v}_{\|} -
\delta\omega)$ in the denominators. The result for the fundamental
$\mathrm{TE}_{10}$~mode has again
form~\eqref{eqn:MicrowaveTrajectories} but now with
\begin{widetext}
\begin{eqnarray}\label{eqn:MicrowaveTrajectories_perp}
r_1 & = & 0, \quad r_0\phi_1\,\, =\,\,
-\frac{\lambda^{3}_{w}|e|\Omega_{0}^{3}C_{0}
     \gamma^{2}_{\|}V_\perp^2}
     {32\pi m_e c^3\gamma_{0}(1 + \bar{v}_{\|}^2\gamma_{\|}^2/c^2)
     \bar{v}^{2}_{\|}}\frac{\omega_{\perp}}{\Omega_0 - \Omega_{\|}}\frac{\exp(i\{[\delta
     k_z\bar{v}_{\|} - \delta\omega]\tilde{t} - \psi\})}
     {(\delta k_z\bar{v}_{\|} -\delta\omega)^{2}}\times \nonumber \\
& &  \hspace{7.1cm}\Bigl\{\frac{\kappa
     + 3\sigma}{\kappa + \sigma} + \frac{k_z^0\lambda_{w}\kappa}{2\pi(\kappa
     + \sigma)} - \frac{\bar{v}_{\|}^2\gamma_0^2\kappa}{4c^2(\kappa
     - \sigma)}\Bigr\}\cos\Bigl(\frac{\pi}{a}\bigl[\bar{x}
     + \frac{V_y}{\Omega_\|}\bigr]\Bigr), \nonumber \\
z_1 & = & \frac{\pi\lambda_{w}|e|\Omega_{0}C_{0}V_\perp}
     {8m_e c\gamma_{0}(1 + \bar{v}_{\|}^{2}\gamma^{2}_{\|}/c^2)
     \bar{v}_{\|}}\frac{\omega_{\perp}}{\Omega_0 - \Omega_{\|}}
     \frac{\exp(i\{[\delta k_z\bar{v}_{\|} - \delta\omega]\tilde{t} - \psi\})}
     {(\delta k_z\bar{v}_{\|} -\delta\omega)^{2}}\times \\
& & \hspace{0cm}\Bigl\{1 +
     \frac{k_z^0\lambda_{w}\kappa}{2\pi(\kappa
     + \sigma)} + \frac{\bar{v}_\|^2\gamma_{0}^2}{4c^2}\Bigl[\frac{\kappa^3
     - 2\kappa^2\sigma - 4\kappa\sigma^2 + 2\sigma^3}{\kappa(\kappa
     - \sigma)(\kappa+2\sigma)} - \frac{2\gamma_{\|}^2(\kappa^3
     + 6\kappa^2\sigma + 8\kappa\sigma^2 - \sigma^3)}{\gamma_0^2\kappa(\kappa + \sigma)
     (\kappa + 2\sigma)}\Bigl]\Bigr\}\cos\Bigl(\frac{\pi}{a}\bigl[\bar{x}
     + \frac{V_y}{\Omega_\|}\bigr]\Bigr), \nonumber
\end{eqnarray}
\end{widetext}
and $\gamma_{\|} = (1 - V_{\|}^2/c^2)^{-1/2}$. It then follows
that the bunching mechanism is substantially axially-azimuthal,
which is in contrast with the undulator synchronism where it is
predominantly axial. In the non-relativistic limit the azimuthal
mechanism does not contribute to the bunching being a
manifestation of dependence of cyclotron frequency on the Lorentz
factor (cf.~\cite{Gaponov1961}). Substituting the obtained result
in~\eqref{eqn:MicrowaveTrajectories} and then
in~\eqref{FO_Eqns_Excitation}, one again gets the dispersion
equation for $\mathrm{TE}_{10}$~mode under the hybrid synchronism
condition.

In this manner, we obtain
\begin{widetext}
\begin{equation}\label{eqn:dispersion_hybrid_1}
\delta k_z (\delta k_z - \frac{\delta\omega}{\bar{v}_{\|}})^2 =
\mp \frac{\pi^{2}|I_0|}{8c\bar{U}\gamma_{0}(1 + \bar{v}_{\|}^2
\gamma_{\|}^2/c^2)}\frac{\omega^2_\perp}{(\Omega_0^2 -
\Omega_{\|}^2)^2}\times \left[
  \begin{array}{r@{,\quad}l@{;}}
  {\displaystyle \frac{\pi V^2_\perp\omega}{a^{3}b\,c}\,Q^{\mathrm{TE}}_{10}\,
  \cos^{2}\Bigl(\frac{\pi}{a}[\bar{x} + \frac{V_y}{\Omega_{\|}}]
  \Bigr)} & \mathrm{TE}_{10}\;\mbox{mode} \\
  {\displaystyle \frac{ V^2_\perp\Omega^2_{0}}{ab^4\Omega^2_{\|}}\,
  Q^{\mathrm{TE}}_{01}\cos^{2}\Bigl(\frac{\pi}{b}[\bar{y} -
  \frac{V_x}{\Omega_{\|}}]\Bigr)} & \mathrm{TE}_{01}\;
  \mbox{mode}
  \end{array}
\right.
\end{equation}
\end{widetext}
which are expansions to the second order in $V_\perp/\bar{v}_{\|}$
of the bulky analytical totally relativistic dispersion equations.
Here the analytically exact expressions for the relativistic
factors read
\begin{widetext}
\begin{eqnarray*}
Q^{\mathrm{TE}}_{10} & = &
      1 - \frac{\bar{v}_{\|}^2\gamma_{\|}^2}{c^2}
      \Bigr\{\frac{8\pi\sigma}{k_z^0\lambda_{w}\kappa}\Bigr[1 -
      \frac{\Omega_{0}(\kappa - \sigma)}{2\omega\kappa}\Bigr] +
      \frac{\Omega_{0}}{2\omega}\Bigr[\frac{(\kappa + \sigma)(\kappa^2
      + \kappa\sigma - \sigma^2)}{\kappa^2(\kappa + 2\sigma)} - \\ 
      & & \hphantom{11 {}- } \frac{\gamma_0^2(\kappa +
      \sigma)(\kappa^3 - 2\kappa^2\sigma - 4\kappa\sigma^2 +
      2\sigma^3)}{2\gamma_{\|}^2(\kappa - \sigma)\kappa^2(\kappa +
      2\sigma)}\Bigr]\Bigr\} + \frac{\pi\bar{v}_{\|}^4\gamma_0^2\Omega_0
      (\kappa + \sigma)\sigma}{c^4 k^0_z\lambda_{w}\omega (\kappa -
      \sigma)\kappa}, \\ 
Q^{\mathrm{TE}}_{01} & = &
      1- \frac{\bar{v}_{\|}k_z^0 b\kappa\sigma}{2c(\kappa^2-\sigma^2)}
      - \frac{2\pi\bar{v}^2_{\|}\gamma_{\|}^2\sigma}{c^2k_z^0\lambda_{w}\kappa}
      + \frac{2\pi\bar{v}^3_{\|}\gamma_0^2 b\sigma^2}{c^3\lambda_{w}(\kappa^2-\sigma^2)}
      - \frac{2\pi^2\bar{v}^5_{\|}\gamma_0^4 b\sigma^3}{c^5 k_z^0
      \lambda_{w}^2\kappa(\kappa^2-\sigma^2)}. 
\end{eqnarray*}
\end{widetext}
Note that $k^{0}_{z} = [(\omega/c)^2-(\pi/a)^2]^{1/2}$ for the
$\mathrm{TE}_{10}$~mode and $k^{0}_{z} = [(\omega/c)^2 -
(\pi/b)^2]^{1/2}$ for the $\mathrm{TE}_{01}$~mode, respectively.

In the non-relativistic limit (substituting $Q^{\mathrm{TE}}_{10}
= 1$ and keeping only the top line of
expression~\eqref{eqn:dispersion_hybrid_1} with $\gamma_{0} =
\gamma_{\|} = 1$ and $\bar{v}_{\|}/c \ll 1$) the obtained
dispersion equation for the fundamental $\mathrm{TE}_{10}$~mode is
similar to that one known in the
literature~\cite[Eq.~(18)]{Marshall1983}. However, even in this
limit, not only the use of Eq.~\eqref{eqn:dispersion_hybrid_1} is
justified by our procedures for all parameter values compatible
with the regular dynamics (i.e. near the magnetoresonance) but it
also provides a good quantitative agreement with the direct
numerical simulations of self-consistent nonlinear
system~\eqref{Master_Eqns} because of the `renormalization'
multipliers $\kappa$ and $\sigma$ (e.g. if one takes $\kappa$ to
be equal to $1$ then $1 + \bar{v}^2_{\|}\gamma^2_{\|}/c^2 \equiv
\gamma^2_{\|})$. It also features analytically calculated
dependence of the right hand side of dispersion equation on the
entrance position $\bar{x}$ of electrons (the factor $\cos^{2}(\pi[\bar{x} +
V_{y}/\Omega_{\|}]/a)$). Thus, one can observe that there exist
two equivalent optimal entrance position of electrons, e.g. for
$|V_{y}|/\Omega_{\|} < a$ the planes $\bar{x} = - V_y/\Omega_{\|}$
(for $V_y < 0$) and $\bar{x} = a - V_y/\Omega_{\|}$  (for $V_y
> 0$), placed near the maxima of gradient of the microwave electric
field  (since the electrons should not touch the waveguide walls
the optimal entrance positions are not always possible). This, in
principle, allows one to double the electron current by using two
electron beams (with the same $x$-components and equal in value
but opposite in direction $y$-components of initial velocity) and
enhance the total spatial growth rate by a quarter (while
simultaneously providing for reduction of aggregate space-charge
effects).

Near the magnetoresonance, analogously to the case of undulator
synchronism, there occurs a substantial reduction in the mean
velocity of constant motion, $\bar{v}_{\|}$, because of kinetic
energy transfer to mainly transversal oscillatory degrees of
freedom. This leads to the gain growth, which is only bounded by
transition to the chaotic state (see
Fig.~\ref{Gain_lambda_hybrid}, $B_{\|}^{res}$ $=$ $3.65$~kG). As
seen in Fig.~\ref{Gain_lambda_hybrid} the magnetoresonance also
influences essentially the dependence of ideal synchronism
($\delta\omega$) frequency on the guide magnetic field. It
should be also emphasized that, as we found in
Sec.~\ref{subsec:NZ_ITV}, the sign of $V_x$ has a substantial
impact on the location of zones of regular and chaotic dynamics,
therefore, this sign can also influence the microwave
amplification. Unfortunately (because the problem involves at least
six parameters), to find a set of paprameters which maximizes
the right hand side of~\eqref{eqn:dispersion_hybrid_1}, we are
bound to confine themselves to detailed numerical calculations.
It turns out that the maximal spatial growth rate in the regular
dynamics zone is attained near the magnetoresonance for $V_x > 0$
and $V_y > 0$, i.e. suppression of dynamical chaos near the
magnetoresonance (achieved by a positive $V_x$, cf.
Fig.~\ref{fig:VelAmpl_Sup/Enh}) may allow one to enhance the
spatial growth rate.

For completeness, we also provide the result of calculations for the
$\mathrm{TE}_{01}$~mode given by the bottom line of
Eq.~\eqref{eqn:dispersion_hybrid_1}, although to implement such an
interaction in a single-mode regime one needs to take additional
efforts for an efficient mode selection (see,
e.g.,~\cite{ArzhannikovEtAl1998}).

\section{\label{sec:NLS} Nonlinear Simulations of Microwave
Amplification}

For numerical simulations of nonlinear regime of amplification it
is advantageous to rewrite equations of motion~\eqref{Eqns_Motion}
taking as independent variables the coordinate~$z$ and time of
entrance of electrons to the interaction region $t_{e}$ (the time
of arrival $t=t(z,t_{e})$ of electrons, which entered the
interaction region at the time $t_{e}$, to the cross-section~$z$
becomes a dependent variable) in the form%
\begin{equation}\label{Eqns_Motion_amplifier}
  \begin{split}
  \frac{d\vec{p}}{dz} & = e\Bigl(\vec{E} +
      \bigl[\frac{\vec{p}}{m_{e}c\gamma}\times
      (\vec{B}_{p} + \vec{B})\bigr]\Bigr)
      \frac{dt}{dz}, \\
  \frac{dt}{dz} & = \frac{m_e\gamma}{p_{z}},
      \quad \frac{d\vec{r}_{\perp}}{dz} =
      \frac{\vec{p}_{\perp}}{p_z},
  \end{split}
\end{equation}
where $\vec{E}(\vec{r}_{\perp},z,t)$,
$\vec{B}(\vec{r}_{\perp},z,t)$ and
$\vec{B}_{p}(\vec{r}_{\perp},z)$ are given by~\eqref{single_mode}
and~\eqref{HybPump_Field}, respectively; $\vec{r}_{\perp} = (x,
y)$; it is also more convenient for numerical calculations to
introduce the momentum of an electron $\vec{p} =
m_{e}\vec{v}\gamma$ instead of its velocity ($\vec{p}_{\perp} =
(p_{x}, p_{y})$, $\gamma = [1 +
\vec{p}\,{}^{2}/(m_{e}^{2}c^{2})]^{1/2}$). The initial conditions
then are $x(z\!\!=\!\!0, t_{e}) = \bar{x}$, $y(z\!\!=\!\!0, t_{e})
= \bar{y}$, $t(z\!\!=\!\!0, t_{e}) = t_{e}$ and
$\vec{p}\,(z\!\!=\!\!0,t_{e}) = m_{e}(V_{x}, V_y,
V_{\|})\gamma_0$. Similarly, using the charge conservation law and
the fact that in the stationary regime electrons, which enter the
interaction region at the time $t_e$ separated by an integral
multiple of the period of the amplified microwave, go along
identical trajectories, we can write the equations of
excitation~\eqref{Eqns_Excitation} for a thin electron beam as
follows~\cite[p.~31]{Kuraev1971}:
\begin{equation}\label{amplitude_exact_num}
\frac{dC}{dz} = \frac{|I_0|}{4\pi P_{0}}\int\limits_0^{2\pi}
p^{-1}_{z}(\vec{p}\cdot\vec{E}^0(\vec{r},t))^{*} \,d(\omega t^e)
\end{equation}
Here the initial condition for the amplitude $C(z)$ reads
$C(z\!\!=\!\!0) = C_{0}$. In the discrete model of electron beam
we assume that over the period of the amplified microwave
particles enter to the interaction region in regular time
intervals $2\pi /(\omega N)$, hence, the numerical finding of
solution to nonlinear self-consistent
system~\eqref{Eqns_Motion_amplifier}
and~\eqref{amplitude_exact_num} consists in the simultaneous
solving of $6N + 1$ first-order nonlinear ordinary differential
equations.

\begin{figure}[t]
\includegraphics[0,0][243,296]{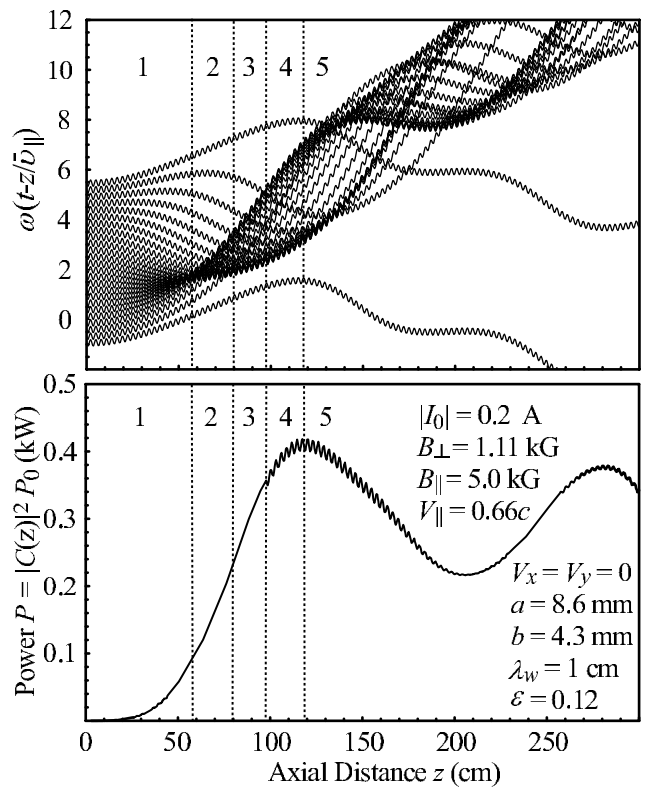}
\caption{\label{Bunch_Hp_1130_beta_066_Hz5000} Results of numerical
calculations of electron beam bunching and microwave power
amplification.}
\end{figure}

Numerical simulations of FEM characteristics are usually concerned
with those, as a rule integral, quantities, which can be directly
measured experimentally (e.g. saturation power, efficiency, start
current, etc.). However, a substantial advantage of numerical
calculations also lies in the opportunity of detailed reconstruction
of electrons dynamics and their interaction with the microwave field
in an FEM. Thus, according to analytical results of
Sec.~\ref{subsec:LTMA_US}, the predominant phasing mechanism on the
undulator mechanism is the axial one. A numerical verification to
this fact is provided in the top of
Fig.~\ref{Bunch_Hp_1130_beta_066_Hz5000}. In the region~1, out of
the uniform at entrance ($z = 0$) electron beam, there occurs bunch
creation under the influence of the seed microwave in its
decelerating phase. The interaction of electron beam with the
microwave is almost linear and the microwave power $P(z) =
|C(z)|^{2}P_{0}$ grows
exponentially with $2\mbox{Im}\,\delta k_z = 0.024~\mbox{cm}^{-1}$.
In the region~2 the electron beam is maximally bunched and
efficiently amplifies the microwave (simultaneously about~$0.35N$ of
the beam electrons, which entered the interaction region at $z = 0$
in the accelerating phase of microwave, draw the power from the
microwave); the total microwave power in this region increases
almost linearly. In the region~3 the electrons, which entered the
interaction region at $z = 0$ in the decelerating phase of
microwave, do not interact with it, but the microwave power growth
takes place because of interaction with approximately~$0.5N$ of beam
electrons that entered the interaction region at $z = 0$ in the
accelerating phase. In the region~4 the energy flow from the
electrons, which amplify the microwave, and those that draw
microwave power reaches the balance and the microwave power attains
saturation. In the region~5 there occurs the second (like the
region~2) consolidation of bunches at $z = 230~\mbox{cm}$, which
subsequently leads to the second maximum of microwave power at $z =
280~\mbox{cm}$ and so on. It is also worth noting that electrons,
which  entered the interaction region at $z = 0$ in the accelerating
phase of the microwave, are weaker trapped than those entering the
interaction region at $z = 0$ in the decelerating phase. %
The small-amplitude fast vibrations shown in
Fig.~\ref{Bunch_Hp_1130_beta_066_Hz5000} are caused by the
oscillations of longitudinal velocity of electrons around its mean
value $\bar{v}_{\|}$ (see~\eqref{ZO_Eqns_Motion_Sols}
and~\eqref{ZO_Eqns_Motion_SolsAdds}). Note that parameter values
used to produce Fig.~\ref{Bunch_Hp_1130_beta_066_Hz5000} are typical
for experimental designs under development, but it is not optimal in
terms of efficiency of microwave amplification because of weak
effective pumping of electron oscillations. As found in
Sec.~\ref{sec:LTMA}, the optimal amplification is achieved near the
magnetoresonance $\Omega_0\approx \Omega_{\|}$. This operational
regime of a hybrid planar FEM is studied in the literature
analytically and numerically in comparatively less detail, mainly,
inasmuch as the major attention of researches has been devoted to
hybrid helix and coaxial FEM schemes. For example, in the hybrid
helix scheme, one uses an annular electron beam, which adiabatic
entrance is incompatible with the magnetoresonant
condition~\cite{Freund1985} since the electrons entering the
interaction region with different radial separations from the
symmetry axis of the undulator magnetic field reach orbits
characterized by a substantial mean velocity spread (this orbits
fail to be the desired stationary helicoidal ones) and do not
provide an adequate amplification of microwaves. In the planar FEM
on the undulator synchronism this drawback is offset substantially
because a sheet electron beam enters in the symmetry plane of the
undulator magnetic field and, therefore, can efficiently amplify
microwaves under the magnetoresonant condition.
\begin{figure}[t]
\centering %
\includegraphics[0,0][244,298]{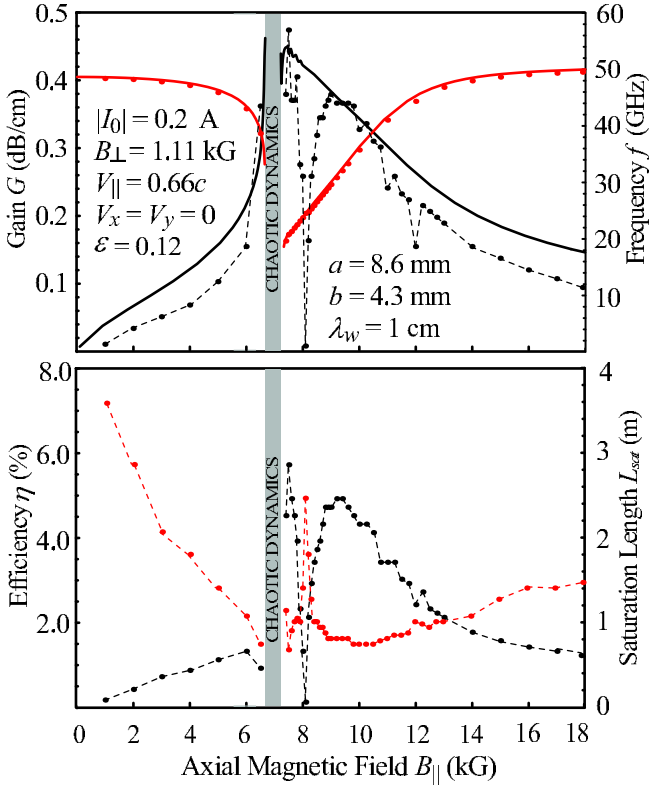}
\caption{\label{wave_Hp_1130_beta_066.eps} In the top figure solid
black (gain $G$) and red (frequency $f$) lines are obtained from
analytical expressions~\eqref{eqn:UndulatorSynch}
and~\eqref{eqn:DispersionUndul}; the dots of respective colors are
the results of direct numerical simulation of
Eqs.~\eqref{Eqns_Motion_amplifier}
and~\eqref{amplitude_exact_num}. In the bottom figure black
(efficiency $\eta$) and red (saturation length $L_{sat}$) dots are
obtained through numerical simulation of
Eqs.~\eqref{Eqns_Motion_amplifier}
and~\eqref{amplitude_exact_num}. All quantities are calculated as
functions of $B_{\|}$ for the $\mathrm{TE}_{10}$~mode
amplification under the undulator synchronism. The darkened area
corresponds to the analytically calculated zones of chaotic
dynamics around the magnetoresonance ($B_{\|}^{res}\approx
6.67~\mbox{kG}$).}
\end{figure}

\begin{figure}[t]
\centering %
\includegraphics[0,0][244,297]{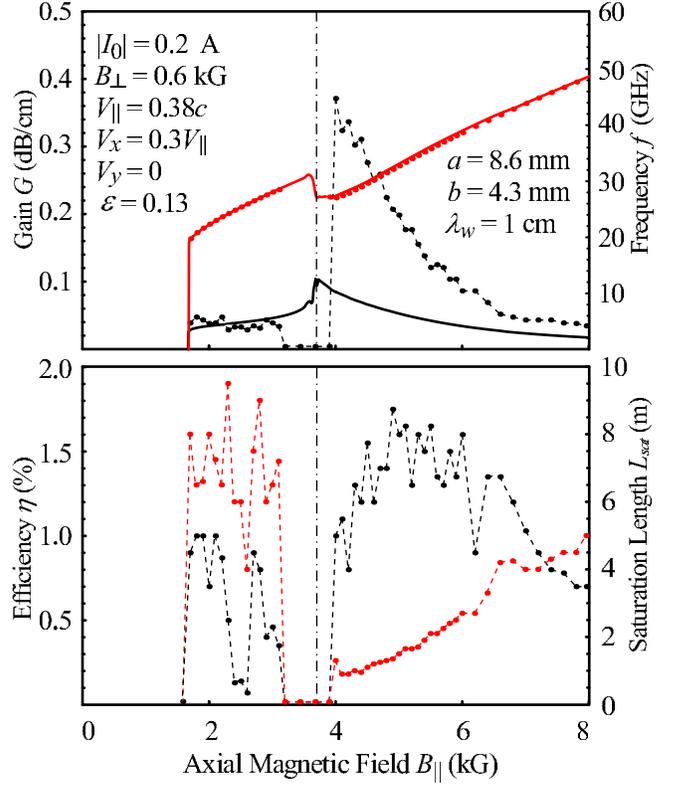}
\caption{\label{Gain_lambda_hybrid} In the top figure solid black
(gain $G$) and red (frequency $f$) lines are obtained from
analytical expressions~\eqref{eqn:HybridSynch}
and~\eqref{eqn:dispersion_hybrid_1}; the dots of respective colors
are the results of direct numerical simulation of
Eqs.~\eqref{Eqns_Motion_amplifier}
and~\eqref{amplitude_exact_num}. In the bottom figure black
(efficiency $\eta$) and red (saturation length $L_{sat}$) dots are
obtained through numerical simulation of
Eqs.~\eqref{Eqns_Motion_amplifier}
and~\eqref{amplitude_exact_num}. All quantities are calculated as
functions of $B_{\|}$ for the $\mathrm{TE}_{10}$~mode
amplification under the hybrid synchronism. The chain lines show
the position of magnetoresonance $B_{\|}^{res}\approx
3.65~\mbox{kG}$ (analytical calculations indicate an exponentially
thin zone of chaotic dynamics for the chosen parameters).}
\end{figure}

Results of numerical calculations of various characteristics of
hybrid planar FEM as functions of the guide magnetic field
$B_{\|}$ are plotted in Figs.~\ref{wave_Hp_1130_beta_066.eps}
and~\ref{Gain_lambda_hybrid}. One can see that the optimal
amplification frequency (the frequency providing for the maximal
growth rate) both for undulator and hybrid synchronisms
(see~\eqref{eqn:UndulatorSynch} and~\eqref{eqn:HybridSynch})
depends on the guide magnetic field since the mean velocity of
constant motion, $\bar{v}_{\|}$, is also a function of $B_{\|}$ at
least through the `renormalization' multiplier $\kappa$. Under the
undulator synchronism the gain, $G$, and efficiency, $\eta$,
attain their maximal values ($G\approx 0.47~\mbox{dB/cm}$ and
$\eta\approx 5.8\%$) for the guide magnetic field ($B_{\|}\sim
7.5~\mbox{kG}$) greater than its magnetoresonant value. A
theoretical estimate of the width of chaotic dynamics zone for the
undulator synchronism gives $B_{\|}\in (6.6, 7.3)~\mbox{kG}$
around the magnetoresonance $B_{\|}^{res}\approx 6.67~\mbox{kG}$.
The situation is similar under the hybrid synchronism with a
slight difference. The gain has a pronounced maximum $G\approx
0.37~\mbox{dB/cm}$ for the guide magnetic fields greater than but
close to the magnetoresonant value $B_{\|}^{res}\approx
3.65~\mbox{kG}$, however, the maximal efficiency $\eta\approx
1.75\%$ is attained at a small finite separation ($B_{\|}\approx
4.9~\mbox{kG}$) from the magnetoresonance. Both cases of undulator
and hybrid synchronisms show a larger efficiency of microwave
amplification for guide magnetic fields greater than their
respective magnetoresonant values since under such conditions a
larger portion of the energy of constant longitudinal motion is
transferred to the transversal oscillations of electrons.

It should be also noted that whereas the growth rate is a slow
function of the amplitude of undulator magnetic field $B_{\perp}$
(or, equivalently, of $\varepsilon$), the efficiency $\eta$ grows
almost linearly with it up to a certain critical value
$\eta_{cr}\approx 20\%$ (e.g.,
$\varepsilon_{cr}\approx 0.2$ under the undulator synchronism). If
$\varepsilon$ exceeds this critical value then the coherent
amplification of microwave signal in the (near) magnetoresonant
regime becomes impossible. In Sec.~\ref{sec:MR} we established
that because of the Nekhoroshev theorem the motion of electrons in
the magnetostatic field of hybrid planar FEM is characterized by
invariant tori, overwhelming majority of which (according to
the Kolmogorov-Arnold-Moser theory, see~\cite{Arnold1993}) under
small perturbations of motion by a microwave signal will not be
destroyed but will only be slightly deformed. Thus, interaction
of electron beam with the microwave signal does not lead to a
notable change of zones of chaotic and regular dynamics if this
interaction is sufficiently small (cf., e.g.,
Figs.~\ref{fig:ContourPlot_ZTV} and \ref{wave_Hp_1130_beta_066.eps}
for the undulator synchronism). However, if the interaction
becomes larger of a certain critical value in the efficiency
(a measure of beam-wave interaction) and, therefore, apparently,
larger of a certain critical value of $\varepsilon$ then such a
notable change of zones of chaotic and regular dynamics does
take place, i.e. there occurs a strong widening of the domain
of disintegrated invariant tori.

\section{Summary and Discussions}

We applied Kisunko-Vainshtein's equations of excitation for regular
waveguides to the development of self-consistent analytical linear
theory of a hybrid planar FEM~amplifier, which is valid not
only far away but also around of the magnetoresonant value of the
guide magnetic field ($\Omega_{\|}\approx \Omega_{0}$). Nonlinear
numerical simulations were undertaken to clarify the validity of
the linear approximation. Although, as mentioned at the
beginning of this article, a number of approaches to the linear
theory of a hybrid FEM has been already formulated, a new one
presented here allowed us both to provide a consistent and unified
analytical description of electron dynamics for all values of the guide
magnetic field and, more importantly, to find an efficient
characterization of the dynamical chaos present in the system.
In particular, conditions of suppression of the dynamical chaos
are formulated and the influence of hybrid planar FEM operational
parameters on these conditions is clarified. It is also worth
mentioning that the zones of suppressed beam transport found
in~\cite{Japanese1993} are in good correspondence with our
(semi-)analytical calculations for the zones of dynamical
chaos (cf. Fig.~1 in~\cite{Japanese1993} and
Figs.~\ref{fig:KappaSigma_Full} and~\ref{fig:ContourPlot_Enh}
of this article). This is achieved because the method of
Lindshtedt employed by us to obtain test individual electron
trajectories in the magnetostatic field of a hybrid planar FEM
is capable of high precision in analytical calculations. It should
be also mentioned that this method is sufficiently general to be
applicable to describe motion of charged particles in
spatially inhomogeneous static magnetic fields like, for example,
that in~\cite{Japanese1993,Destler1996}.

The origin of dynamical chaos is connected to the possibility of
onset of stochastic layer around the separatrix in the system phase space
(under the condition $\omega_{\perp}/\omega_{0} = [(1 + V^{2}_{x}/V^{2}_{\|})^{1/2}
+ V_{x}/V_{\|}]/2$ motion of an electron in the undulator magnetic field
takes place along the separatrix) caused by the presence of guide magnetic field,
which plays the role of non-trivial perturbation capable of destruction
of this separatrix (see Figs.~\ref{fig:ContourPlot_ZTV},~\ref{fig:ContourPlot_Sup}
and~\ref{fig:ContourPlot_Enh}). The magnitude and sign of the $x$-component of
initial velocity of electrons influence the position of separatrix in the
phase space and allows one to adjust location of zones of regular and
chaotic dynamics when the guide magnetic field is present. An approximate
condition that determines the onset of chaos is given analytically and
found to be in a good quantitative agreement with numerical simulations.
We interpret this condition as the Chirikov resonance-overlap criterion,
i.e. the chaotic behavior occurs whenever the absolute value of the
difference between the normal undulator,~$\Omega_{0}$, and normal
cyclotron,~$\Omega_{\|}$, frequencies becomes less than the
coupling,~$\omega_{\perp}$, induced by the undulator
magnetic field. It seems also that the analytical approach
developed in this paper has a strong potential for a development
of analytical nonlinear in the microwave signal three-dimensional
theory of a hybrid FEM.

The transfer of kinetic energy of an electron motion between
longitudinal and transversal degrees of freedom is studied
and it is found that the maximal fraction,~$(\bar{\gamma}_{\perp} -
1)/(\gamma - 1)$, of the mean transversal kinetic energy
is less than 30\% for the case of regular dynamics. Taking into
account that transversal degrees of freedom of electrons are the
source of energy for the microwave field one can claim
that the maximal efficiency of a hybrid planar FEM may
not exceed these 30\%. Using the Nehoroshev theorem we showed that
motion of an individual test electron in the magnetostatic field is
integrable and characterized by two-dimensional invariant tori for
some range of parameters of hybrid planar FEM. As a result from the
Kolmogorov-Arnold-Moser theory it then follows that for
relatively moderate values of the microwave and space-charge
field the electron trajectories will stay regular and amplification
can still be accomplished. Through numerical simulations, we
determined that the broadening of chaotic region for electron
motion does not occur if the interaction between the
microwave and electron beam is not too large (e.g., if such a
measure of interaction intensity as the efficiency is less
than 20\% on the undulator synchronism).

From linearized in the microwave field equations of motion and
excitation dispersion equations are derived for the undulator
and hybrid synchronisms. We showed analytically that around the
magnetoresonance on the
undulator synchronism the gain is nearly completely independent
of the amplitude of the undulator magnetic field. This
circumstance is known in the literature but only as a result of
numerical simulations (cf. Fig.~2 in~\cite{NGorky2001}). Physical
origin of this effect lies in the fact that the gain is a function
of amplitude of transversal oscillations of electrons in the
magnetostatic field of a hybrid planar FEM. Around the magnetoresonant
value of guide magnetic field this amplitude, according to the
chaotization criterium~\eqref{eqn:ChirikovCriterium}, turns out
to be bounded and is independent of the amplitude of undulator
magnetic field. On the undulator synchronism the obtained
analytical expression for the gain provides values close
to the results of nonlinear numerical simulations. Such an accuracy
is achieved because of
the high-precision analytical calculation of the unperturbed by
the microwave proper frequencies and electron trajectories. For
the hybrid synchronism we also obtained the dispersion equation
but in this case the analytical expression for the gain in the
zone of regular electron dynamics immediately below and above the
magnetoresonant value of the guide magnetic field does not provide
values close to the results of numerical calculations. We attribute
this discrepancy to a greater amplitude of oscillations of the
longitudinal electron velocity and, therefore, to a greater
deviation (comparing to the case of the undulator synchronism)
of the resultant system of ordinary differential
equations from that with the strictly periodic coefficients. Solutions
to such ordinary differential equations with the quasi-periodic
coefficients does not follow patterns of their counterparts with
the periodic coefficients especially in the vicinity of (combination)
resonances between their proper frequencies. Nevertheless, we showed
analytically and verified through numerical simulations that operation
of a hybrid planar FEM in the (near) magnetoresonant regime is optimal
in order to achieve the maximal gain and efficiency. The interaction
between transversal degrees of freedom of electrons and microwave
field under such conditions turns out to be maximal out of all other
possible operation parameters of a hybrid planar FEM on each
of the synchronisms (experimentally for a hybrid planar FEM~oscillator
such an operational regime was established, for example,
in~\cite{NGorky1996}). The major result here is that one needs not only
to have a small value of $\varepsilon$ in order to maintain the relation
$\bar{v}_{\|}\gg \bar{v}_{\perp}$ ($\bar{v}_{\perp}$ is the mean
transversal velocity) but also to hold the ratio
$\varepsilon /|\kappa - \sigma|\equiv \omega_{\perp}/|\Omega_{0} -
\Omega_{\|}|$ as close to the unity as possible thus providing for the
maximal gain and efficiency.

Here we have not considered any specific arrangements usually applied
for a smooth (at best adiabatic) entrance of an electron beam to the
interaction region, which lead us to neglect of the velocity and
position spread of electrons at the entrance to the interaction region.
The developed techniques are fully capable of treating these effects
but taking them into account in the framework of the current
consideration would make our treatment overcomplicated and hide behind
the technicalities the physics underlying the nature of dynamical chaos in
a hybrid planar FEM. This paper also does not deal with the influence
of the space-charge field on operational characteristics of a hybrid
planar FEM. It is necessary to notice that space-charge field strongly
decreases the efficiency of interaction between an electron beam and
amplified microwave if the regime of operation is positioned far away
from the magnetoresonance. However, there are strong indications that not
only the amplification of a microwave signal by an electron beam
becomes maximal around the magnetoresonance but also the defocusing
influence of the present space-charge field turns out to be minimal.
A detailed investigation of the influence of potential (irrotational)
and rotational parts of the space-charge field on the operation of a
weakly-relativistic hybrid FEM will be published elsewhere.

\begin{acknowledgments}
We acknowledge fruitful conversations and discussions with
V.L.~Bratman, N.S.~Ginzburg, N.Yu.~Peskov, O.V.~Usatenko,
and V.V.~Yanovsky.
\end{acknowledgments}

\appendix*
\section{Velocity to the Order $o(\varepsilon^3)$}\label{App:Velocity}

Here we present terms of the order $\mathcal{O}(\varepsilon^{2})$,
which should be added componentwise to
expressions~\eqref{ZO_Eqns_Motion_Sols} to obtain solutions to the
electron velocities valid to the order $o(\varepsilon^3)$. These
additional contributions read
\begin{widetext}
\begin{eqnarray}\label{ZO_Eqns_Motion_SolsAdds}
        \dot{x}_{0}^{add}(\tilde{t}) & = &
-2V_{\perp}\frac{\omega_{\perp}^{2}\Omega_{0}^{3}
\Omega_{\|}}{(\Omega_{0}^{2} - \Omega_{\|}^{2})^{3}}
\Bigl[\sin(\Omega_{0}\tilde{t})\sin\psi - \frac{3\Omega_{0}^{4} +
6\Omega_{0}^{2}\Omega_{\|}^{2} -
\Omega_{\|}^{4}}{16\Omega_{0}^{3}\Omega_{\|}}
\cos(\Omega_{\|}\tilde{t} - \psi) - \frac{\Omega_{0}^{2} -
5\Omega_{\|}^{2}}{16\Omega_{0}\Omega_{\|}}
\cos(\Omega_{\|}\tilde{t} + \psi) + \nonumber \\
        &  & \frac{(\Omega_{0} - \Omega_{\|})^{3}(2\Omega_{0} +
\Omega_{\|})^{2}}{32\Omega_{0}^{4}\Omega_{\|}} \cos([2\Omega_{0} +
\Omega_{\|}]\tilde{t} - \psi) + \frac{(\Omega_{0} +
\Omega_{\|})^{3}(2\Omega_{0} -
\Omega_{\|})^{2}}{32\Omega_{0}^{4}\Omega_{\|}}
\cos([2\Omega_{0} - \Omega_{\|}]\tilde{t} + \psi)\Bigr], \nonumber \\
        \dot{y}_{0}^{add}(\tilde{t}) & = &
-2V_{\perp}\frac{\omega_{\perp}^{2}\Omega_{0}^{2}
\Omega_{\|}^{2}}{(\Omega_{0}^{2} - \Omega_{\|}^{2})^{3}} \Bigl[
\cos(\Omega_{0}\tilde{t})\sin\psi + \frac{3\Omega_{0}^{4} +
6\Omega_{0}^{2}\Omega_{\|}^{2} -
\Omega_{\|}^{4}}{16\Omega_{0}^{2}\Omega_{\|}^{2}}
\sin(\Omega_{\|}\tilde{t} - \psi) + \frac{\Omega_{0}^{2} -
5\Omega_{\|}^{2}}{16\Omega_{\|}^{2}}
\sin(\Omega_{\|}\tilde{t} + \psi) - \nonumber \\
        &  & \hspace{-4em}\frac{(\Omega_{0} + \Omega_{\|})^{3}(2\Omega_{0}
- \Omega_{\|})}{32\Omega_{0}^{3}\Omega_{\|}} \sin([2\Omega_{0} -
\Omega_{\|}]\tilde{t} + \psi) - \frac{(\Omega_{0} -
\Omega_{\|})^{3}(2\Omega_{0} +
\Omega_{\|})}{32\Omega_{0}^{3}\Omega_{\|}}\sin([2\Omega_{0} +
\Omega_{\|}]\tilde{t} - \psi) + \frac{\Omega_{0}^{4} -
\Omega_{\|}^{4}}{8\Omega_{0}^{2}\Omega_{\|}^{2}}\sin\psi\Bigr], \nonumber \\
        \dot{z}_{0}^{add}(\tilde{t}) & = &
-\frac{\omega_{\perp}^{2}\bar{v}_{\|}^{6}\Omega_{0}\lambda_{w}}{32\pi
V_{\|}^{6}(\Omega_{0}^{2} - \Omega_{\|}^{2})}\Bigl[\Bigl\{4 -
2\frac{V_{\perp}^{2}(\Omega_{0}^{2} +
\Omega_{\|}^{2})}{\bar{v}_{\|}^{2}(\Omega_{0}^{2} -
\Omega_{\|}^{2})}\Bigr\}\cos(2\Omega_{0}\tilde{t}) +
4\frac{V_{\perp}^{2}\Omega_{0}^{2}}{\bar{v}_{\|}^{2}(\Omega_{0}^{2}
- \Omega_{\|}^{2})}\cos(2\Omega_{\|}\tilde{t} - 2\psi) - \\
       &  & 8\frac{\Omega_{0}^{3}(\Omega_{0} +
\Omega_{\|})}{(\Omega_{0}^{2} - \Omega_{\|}^{2})^{2}}\Bigl\{\Bigl(1
- \frac{\Omega_{\|}^{2}}{\Omega_{0}^{2}} +
\frac{V_{\perp}^{2}\Omega_{\|}}{\bar{v}_{\|}^{2}\Omega_{0}}\Bigr)\cos([\Omega_{0}
- \Omega_{\|}]\tilde{t}) -
\frac{V_{\perp}^{2}\Omega_{\|}}{\bar{v}_{\|}^{2}\Omega_{0}}\cos([\Omega_{0}
- \Omega_{\|}]\tilde{t} + 2\psi)\Bigr\} - \nonumber 
\end{eqnarray}
\begin{eqnarray}
       &  & 8\frac{\Omega_{0}^{3}(\Omega_{0} -
\Omega_{\|})}{(\Omega_{0}^{2} -
\Omega_{\|}^{2})^{2}}\Bigl\{\Bigl(1 -
\frac{\Omega_{\|}^{2}}{\Omega_{0}^{2}} -
\frac{V_{\perp}^{2}\Omega_{\|}}{\bar{v}_{\|}^{2}\Omega_{0}}\Bigr)\cos([\Omega_{0}
+ \Omega_{\|}]\tilde{t}) +
\frac{V_{\perp}^{2}\Omega_{\|}}{\bar{v}_{\|}^{2}\Omega_{0}}\cos([\Omega_{0}
+ \Omega_{\|}]\tilde{t} - 2\psi)\Bigr\} - \nonumber \\
&  & \frac{V_{\perp}^{2}}{\bar{v}_{\|}^{2}}\Bigl\{
\frac{\Omega_{0}(\Omega_{0} + \Omega_{\|})}{(\Omega_{0} -
\Omega_{\|})^{2}}\cos(2[\Omega_{0} - \Omega_{\|}]\tilde{t} +
2\psi) + \frac{\Omega_{0}(\Omega_{0} - \Omega_{\|})}{(\Omega_{0} +
\Omega_{\|})^{2}}\cos(2[\Omega_{0} + \Omega_{\|}]\tilde{t} -
2\psi)\Bigr\} \Bigr]. \nonumber
\end{eqnarray}
\end{widetext}

\newpage 
\bibliography{draft}

\end{document}